\documentclass[11pt]{article}

\usepackage[T1]{fontenc}
\usepackage{amsfonts, amsmath, amssymb}
\usepackage{mathtools}
\usepackage{mathrsfs}
\usepackage[a4paper, left=2.5cm, right=2cm, top=2.5cm, bottom=2.5cm]{geometry}
\usepackage{graphicx}
	\graphicspath{{ext/}, {plt/}, {plt/ftx/}} 
\usepackage[dvipsnames]{xcolor} 
\usepackage{tikz}
	\usetikzlibrary{external, positioning, arrows.meta, calc, decorations.markings}
	\tikzexternaldisable{} 
\usepackage{pgfplots}
	\pgfplotsset{compat=newest} 
	\usepgfplotslibrary{colorbrewer, statistics, groupplots, fillbetween}
	\newlength{\figurewidth}
	\newlength{\figureheight}
\usepackage{pgfplotstable}
\usepackage{colortbl} 
\usepackage{booktabs}
\usepackage{multirow}
\usepackage[version=4]{mhchem} 
\usepackage{algorithm}
\usepackage{algorithmicx}
\usepackage{algpseudocode}
	\algdef{S}[FOR]{Foreach}[1]{\algorithmicforeach\ #1\ \algorithmicdo}
	\algrenewcommand{\textproc}{\normalfont}
	\algrenewcommand\algorithmiccomment[1]{\hfill \textcolor{Cyan}{\texttt{//\;#1}}}
	\algnewcommand\algorithmicinput{\textbf{Input: }}
	\algnewcommand\algorithmicoutput{\textbf{Output: }}
	\algnewcommand\algorithmicforeach{\textbf{for each}}
	\algdef{S}[FOR]{ForEach}[1]{\algorithmicforeach\ #1\ \algorithmicdo}
	\algnewcommand\algorithmictimes{\textbf{times}}
	\algdef{E}[REPEAT]{Times}[1]{#1\ \algorithmictimes}
	\algrenewcommand\textproc{\textsf}
	\algnewcommand{\IfThen}[2]{\State \algorithmicif\ #1\ \algorithmicthen\ #2}
\usepackage[colorlinks=true]{hyperref} 
\hypersetup{
    linkcolor=RedViolet,
    urlcolor=OliveGreen,
    citecolor=PineGreen
}
\usepackage[capitalise, noabbrev, nameinlink]{cleveref}
	\crefname{equation}{equation}{equations} 
\usepackage{autonum} 
\usepackage{tcolorbox} 
    \tcbuselibrary{most}
\usepackage{siunitx}
\usepackage{lipsum}
\usepackage{tabularx}
\usepackage{environ} 
\usepackage{authblk}
\usepackage[title]{appendix}

\newcommand{\bsc}{{\boldsymbol{c}}}

\newcommand{\bse}{{\boldsymbol{e}}}

\newcommand{\bsq}{{\boldsymbol{q}}}

\newcommand{\bsu}{{\boldsymbol{u}}}

\newcommand{\bsx}{{\boldsymbol{x}}}
\newcommand{\bsy}{{\boldsymbol{y}}}
\newcommand{\bsz}{{\boldsymbol{z}}}

\newcommand{\bsP}{{\boldsymbol{P}}}

\newcommand{\bszero}{{\boldsymbol{0}}}

\newcommand{\bsxi}{{\boldsymbol{\xi}}}

\newcommand{\bbE}{{\mathbb{E}}}

\newcommand{\bbV}{{\mathbb{V}}}

\newcommand{\N}{{\mathbb{N}}} 

\DeclareSymbolFont{bbold}{U}{bbold}{m}{n}
\DeclareSymbolFontAlphabet{\mathbbold}{bbold}

\newcommand{\calI}{{\mathcal{I}}}

\newcommand{\calL}{{\mathcal{L}}}

\newcommand{\fraku}{{\mathfrak{u}}}

\DeclareMathOperator{\E}{\bbE}
\DeclareMathOperator{\V}{\bbV}

\providecommand{\argmin}{\operatorname*{argmin}}
\providecommand{\argmax}{\operatorname*{argmax}}
\DeclareMathSymbol{\shortminus}{\mathbin}{AMSa}{"39}

\tikzset{%
	>={Stealth[length=2mm, width=1.75mm]}, 
	default line/.style={%
		thick,
		line cap=round,
	},
	default dashed line/.style={%
		default line,
		dashed,
	},
	default marker line/.style={%
		default line,
		mark=*,
		mark size=0.75pt,
	},
}

\pgfkeys{%
	/pgf/number format/.cd,1000 sep={}
}

\pgfplotsset{%
	default axis/.style={%
		width=\figurewidth,
		height=\figureheight,
		major tick length={2pt},
		minor tick length={2pt},
		every tick/.style={black, line cap=round},
		ticklabel style={font=\scriptsize},
		x label style={font=\small},
		y label style={font=\small},
		legend style={%
			draw=none,
			font=\scriptsize,
			at={(1.03, 1)},
			anchor=north west,
			fill=none,
			legend cell align=left
		},
		cycle list/Set1,
	},
	default error/.style={%
		error bars/.cd,
		y dir=both,
		y explicit,
	}
}

\tcbset{%
	outer arc=2pt,
	arc=1pt,
	attach boxed title to top left={yshift=-3mm, xshift=3mm},
	enhanced,
	boxed title style={%
		top=3pt,
		bottom=3pt
	},
	fonttitle=\sffamily,
	top=12pt
}

\newtcolorbox[auto counter, number within=section]{todobox}[1][]{%
	colframe=NavyBlue,
	colback=NavyBlue!20,
	boxed title style={%
		colback=NavyBlue
	},
  	title=TO DO,
}

\newtcolorbox[auto counter, number within=section]{infobox}[1][]{%
	colframe=OliveGreen,
	colback=OliveGreen!20,
	boxed title style={%
		colback=OliveGreen
	},
	title=INFO,
}

\newtcolorbox[auto counter, number within=section]{warningbox}[1][]{%
	colframe=Red,
	colback=Red!20,
	boxed title style={%
		colback=Red
	},
  	title=WARNING,
}

\newcommand{\FTridyn}{\texttt{F-TRIDYN}}
\newcommand{\Xolotl}{\texttt{Xolotl}}
\newcommand{\FTX}{\texttt{FTX}}
\newcommand{\GITR}{\texttt{GITR}}
\newcommand{\ITERHe}{\textsf{ITER-He}}
\newcommand{\PISCES}{\textsf{PISCES-A}}

\newcommand{\plotsens}[4]{%
    \pgfmathtruncatemacro{\bottom}{#1 - 1}
    \pgfmathtruncatemacro{\top}{#1}
    \addplot[name path=bottom, draw=none, smooth, forget plot] table[x index=0, y index=\bottom] {#3};
    \addplot[name path=top, draw=none, smooth, forget plot] table[x index=0, y index=\top] {#3};
    \addplot+[%
        draw=none,
        color=#2,
        legend image code/.code={\expandafter\draw [yshift=-0.4ex] (0ex,0ex) rectangle (1ex,1ex);},
        #4
    ] fill between[of=bottom and top];
}

\newcommand{\inputbullet}{\textbullet\;}
\newcommand{\outputbullet}{$\clubsuit$\;}
\newcommand{\transferbullet}{$\spadesuit$\;}

\title{Global Sensitivity Analysis of a Coupled Multiphysics Model to Predict Surface Evolution in Fusion Plasma-Surface Interactions}

\author[1]{Pieterjan Robbe}
\author[2]{Sophie Blondel}
\author[1]{Tiernan Casey}
\author[2]{Ane Lasa}
\author[1]{\authorcr Khachik Sargsyan}
\author[2,3]{Brian D. Wirth}
\author[1]{Habib N. Najm}

\affil[1]{Sandia National Laboratories, Livermore, CA 94551, USA}
\affil[2]{University of Tennessee, Knoxville, TN 37996, USA}
\affil[3]{Oak Ridge National Laboratory, Oak Ridge, TN 37831, USA}

\begin{document}

\maketitle

\begin{abstract}
    We construct a global sensitivity analysis framework for a coupled multiphysics model used to predict the changes in material properties and surface morphology of helium plasma-facing components in future fusion reactors. The model combines the particle dynamics simulator \FTridyn{}, that predicts the helium implantation profile, with the cluster dynamics simulator \Xolotl{}, that predicts the growth and evolution of subsurface helium gas bubbles. In order to keep the sensitivity analysis tractable, we first construct a sparse, high-dimensional polynomial chaos expansion surrogate model for each output quantity of interest, which allows the efficient extraction of sensitivity information. The sensitivity analysis is performed for two problem settings: one for ITER-like conditions, and one that resembles conditions inside the PISCES-A linear plasma device. We present a systematic comparison of important parameters, for both \FTridyn{} and \Xolotl{} in isolation as well as for the coupled model, and discuss the physical interpretation of these results.
\end{abstract}

\section{Introduction}\label{sec:introduction}

Nuclear fusion is a potential source of safe, carbon-free and virtually limitless energy. However, the design of next-generation tokamak fusion reactors like ITER and DEMO is challenged by difficulties in understanding and predicting material damage in plasma-facing components (PFCs)~\cite{meade2009}. Tungsten (\ce{W}) is a premier candidate for the reactor wall material, because it can withstand the harsh thermal and radiation loads present in these reactors. Unfortunately, its material properties degrade when exposed to the fusion plasma, causing detrimental effects on the operation of the fusion reactor~\cite{brezinsek2015}. Experiments with linear plasma devices, as well as in situ tokamak plasmas, have shown that the helium (\ce{He}) generated by the fusion reaction causes drastic changes in the microstructure of the tungsten wall, leading to erosion, material performance degradation and ultimately material failure~\cite{hammond2017,wright2013}.

The need for fundamental insight into the mechanics of the microstructure evolution in plasma-surface interactions (PSI) has motivated a variety of computational studies and modeling attempts. For example, in~\cite{sefta2013}, it was shown that helium atoms quickly form clusters that grow inside the tungsten lattice. Larger helium clusters diffuse below the tungsten surface, until they eventually become over-pressurized and burst. The presence of these subsurface gas bubbles produce high levels of stress in the tungsten matrix, causing a change in the mechanical properties of the surface material and the surface morphology. The evolution of the subsurface gas inventories at short time scales has been investigated in~\cite{sefta2013a, ito2014, hammond2018}. In these studies, Molecular Dynamics (MD) simulations, which model the physical movements of individual helium atoms in the tungsten lattice, are used as the main computational technique. Because of computational constraints, and despite acceleration attempts proposed in~\cite{sandoval2015}, these MD simulations are typically restricted to small spatial domains, large implantation fluxes and short timescales, conditions that are far away from those associated with the available experimental data.

One way to overcome these challenges is to use a continuum-scale reaction-diffusion cluster dynamics model, see, e.g.,~\cite{blondel2017, perez2017, blondel2018}. In this model, microstructure evolution is described in terms of concentration fields of helium and vacancy clusters, as opposed to the modeling of individual atoms in MD. In this work, we will use the cluster dynamics software \Xolotl{}~\cite{xolotl}. It has been shown that \Xolotl{} captures the essential physical behavior governing the tungsten subsurface evolution, and quantitatively reproduces the values predicted by representative MD simulation benchmarks~\cite{blondel2017, blondel2017a, maroudas2016}.

\Xolotl{} has been combined with the ion-solid interaction software \FTridyn{} in~\cite{lasa2020}. \FTridyn{} is a particle code that predicts the helium implantation profile and sputtering yields of helium and tungsten, that are used as inputs in \Xolotl{}~\cite{drobny2017}. In what follows, we will denote coupled \FTridyn{}-\Xolotl{} as \FTX{}.

The model implemented in \FTX{} depends on 17 input parameters whose values have been obtained from accompanying MD simulations. However, a significant uncertainty is associated with each of these parameter values, resulting in a corresponding uncertainty in the predicted model outcomes. To increase the fidelity of the model predictions, it is important to improve understanding about the global sensitivity of model outputs with respect to specific uncertain parameters, as well as the contribution of parametric uncertainty to the overall predictive capability of the model.

Variance-based Global Sensitivity Analysis (GSA), involving the computation of Sobol' indices, is particularly attractive, because it allows one to attribute fractions of the model output variance to the variability in the input parameters~\cite{saltelli2008, sobol2001}. However, this approach typically involves the computation of conditional expectations, and, when these are computed using random sampling, requires a prohibitively large number of model evaluations to reach an acceptable level of accuracy. An alternate approach to obtain the Sobol' indices is to first construct a surrogate for the model response using a limited set of training samples, and to replace the expensive model with the surrogate when evaluating the integrals in the expectation. Of particular interest are polynomial chaos expansion (PCE) surrogate models~\cite{ghanem1991,ghanem1999,lemaitre2001,reagan2003,najm2009,ernst2012}, because they allow estimation of sensitivities directly from the PCE~\cite{sudret2008, crestaux2009}.

In this paper, we develop a GSA approach for \FTX{}, focusing on the prediction of surface evolution and helium retention. The latter is the amount of subsurface helium retained in the tungsten material during exposure to the fusion plasma, and is an important performance metric of PFCs. Our framework is based on PCE surrogate-based sensitivity analysis, where we exploit recent advancements in sparse and adaptive constructions~\cite{sargsyan2014}. We use our framework to perform sensitivity analysis in two problem settings:
\begin{itemize}
    \item \ITERHe{}, a problem setting for ITER-like conditions using a helium plasma~\cite{lasa2020}, and
    \item \PISCES{}, a problem setting that mimics the PISCES-A linear plasma device~\cite{baldwin2010, doerner2001, baldwin2019}.
\end{itemize}
A similar sensitivity analysis for predicting the impact of plasma impurities in the \PISCES{} setting using the impurity transport code \GITR{}, see~\cite{younkin2021}, has been performed in~\cite{younkin2022}. However, no sensitivity analysis framework has yet been proposed for the formation of subsurface gas bubbles in tungsten materials exposed to helium plasma. This will be the focus of the present work.

The structure of this paper is as follows. First, in~\cref{sec:multiphysics_modeling_of_plasma-exposed_surfaces}, we give a brief overview of \FTridyn{} and \Xolotl{}, the two codes used to predict the surface evolution of PFCs under exposure to helium plasmas. Next, in \cref{sec:global_sensitivity_analysis_framework}, we discuss our approach for sensitivity analysis, including the construction of PCE surrogate models. In \cref{sec:results_and_discussion}, we present our main results for both the \ITERHe{} (\cref{sec:gsa_study_for_the_iter_he_setting}) and \PISCES{} (\cref{sec:gsa_study_for_the_pisces_setting}) problem settings, followed by a discussion in \cref{sec:discussion}. Conclusions and pointers to future work are presented in \cref{sec:conclusions_and_future_work}.

\section{Multiphysics modeling of plasma-exposed surfaces}\label{sec:multiphysics_modeling_of_plasma-exposed_surfaces}

\FTX{} is part of a larger multiphysics modeling framework for predicting the evolution of PFCs under irradiation, including physical phenomena spanning multiple decades in length and time scales, such as clustering and bursting of helium bubbles, tungsten wall material erosion, transport and redeposition of eroded impurities, and the evolution of the scrape-off layer plasma inside the reactor~\cite{lasa2020, lasa2021}. Many of these physical processes are strongly coupled, since the plasma response is strongly dependent on the material surface evolution and vice versa.

The multiscale framework from~\cite{lasa2020, lasa2021} involves a hierarchy of multiphysics codes, including \texttt{SOLPS} for the simulation of the steady-state background plasma~\cite{schneider2006}, \texttt{hPIC} to resolve the near-surface sheath effects~\cite{khaziev2015}, \texttt{GITR} to calculate the migration and redeposition of impurities eroded from the tungsten surface~\cite{younkin2021}, and finally \FTX{} to compute the sputtering and implantation of ions impacting the material and to obtain the response of the wall surface under these plasma conditions~\cite{lasa2020, lasa2021}.

\begin{figure*}[t]
    \centering
    \small
    \let\nodehdist\relax\newlength\nodehdist
\let\nodevdist\relax\newlength\nodevdist
\setlength{\nodehdist}{2cm}
\setlength{\nodevdist}{0.75cm}
\begin{tikzpicture}[%
        node distance=\nodevdist and \nodehdist,
        box/.style={%
            text width=3cm,
            align=center,
            draw,
            inner sep=5pt,
            execute at end node={\strut},
            minimum height=1.5cm
        },
    ]
    \node[box] (ftridyn) {\FTridyn{}};
    \node[box, below right=of ftridyn] (xolotl) {\Xolotl{}};
    \draw[->] (ftridyn) -| node[pos=0, anchor=south west, text width=5cm, inner sep=6pt] {\transferbullet{}\strut sputtering yields \ce{W} / \ce{He} \\ \transferbullet{}\strut implantation profile} (xolotl.150);
    \draw[->] (xolotl) -| node[pos=0, anchor=north east, text width=3.75cm, inner sep=6pt] {\transferbullet{}\strut material composition} (ftridyn);
    \draw[->] ([xshift=-1cm, yshift=0.85cm]ftridyn.west) |- node[pos=0, anchor=193, text width=9.5cm, inner sep=6pt] {\inputbullet{}\strut Ion Energy-Angle Distribution (IEAD) \\ \inputbullet{}\strut cutoff energies of \ce{W} and \ce{He} ($E_{f, \ce{W}}$ and $E_{f, \ce{He}}$) \\ \inputbullet{}\strut tungsten surface binding energy ($E_s$)} (ftridyn.west);
    \draw[->] (ftridyn -| xolotl.30) -- node[pos=0, anchor=205, text width=6cm, inner sep=6pt, xshift=1em] {\inputbullet{}\strut lattice parameter ($a_0$) \\ \inputbullet{}\strut \ce{He_1} radius ($r_{\ce{He}_1}$) \\ \inputbullet{}\strut interstitial bias factor ($b_{\ce{I}_l}$) \\ \inputbullet{}\strut initial vacancy concentration ($C_{\ce{V}_1}$) \\ \inputbullet{}\strut migration energies ($E_{\ce{He}_m}$ and $E_{\ce{V}_1}$)} (xolotl.30);
    \draw[->] (xolotl.east) -- node[pos=0, anchor=north west, text width=4cm, inner sep=6pt] {\outputbullet{}\strut helium retention \\ \outputbullet{}\strut surface growth} ([xshift=1cm]xolotl.east);
\end{tikzpicture}
    \caption{Schematic overview of the coupling between \FTridyn{} and \Xolotl{}, exposing the relevant uncertain input parameters and the relevant model outputs. Input parameters are denoted by \inputbullet{}, outputs are denoted by \outputbullet{}, and information exchanged between \FTridyn{} and \Xolotl{} is denoted by \transferbullet{}. See also \cref{tab:uncertain_parameters} for an overview of each parameter.}
    \label{fig:coupling_ftridyn_xolotl}
\end{figure*}
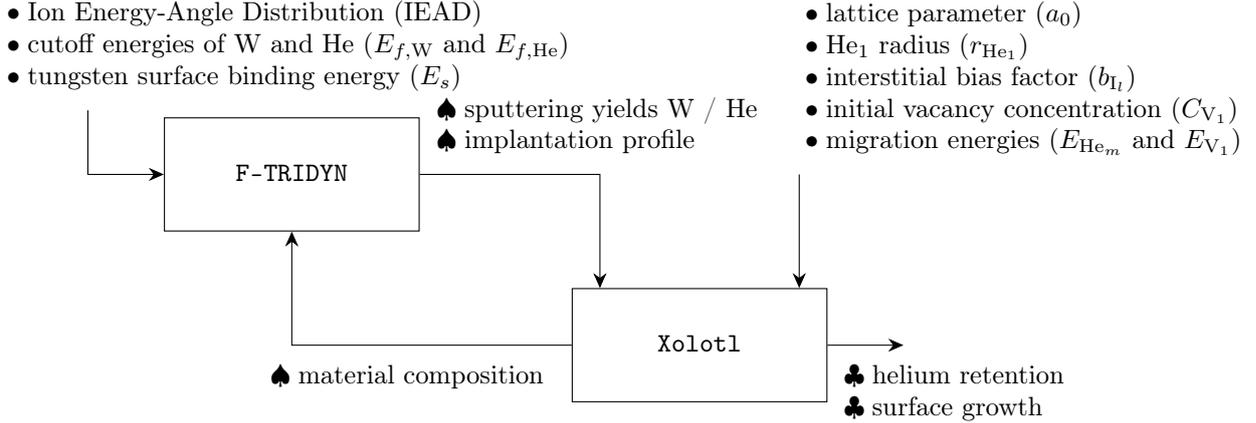

In the coupled \FTX{} setting, the ion energy-angle distribution (IEAD), describing the distribution of the incoming kinetic energy as a function of the impact angle for each ion species in the plasma and computed by \texttt{hPIC}, is used as an input to \FTridyn{}. \FTridyn{} provides the implantation profiles of the incident helium plasma ions, i.e., the concentration of helium ions implanted in the material as a function of depth, as well as the sputtering yields of helium and tungsten, i.e., the average number of atoms removed from a target per incident helium particle, to \Xolotl{}. \Xolotl{} in turn then models the subsurface material composition as a function of time, based on the diffusion and reaction of subsurface helium atoms and defects. The time-dependent simulation of \FTX{} involves running \FTridyn{} and \Xolotl{} in tandem within each time step, as the subsurface composition predicted by \Xolotl{} affects the implantation profile and sputtering yield predicted by \FTridyn{} and vice versa. We refer to~\cite{lasa2021} for more details regarding the different aspects of the code coupling. A schematic overview of the \FTX{} coupling is shown in \cref{fig:coupling_ftridyn_xolotl}. We remark that the \FTX{} workflow also uses particle and heat fluxes from \texttt{SOLPS}, the IEAD of tungsten particles from \texttt{hPIC}, as well as the plasma composition predicted by \texttt{GITR}. However, these dependencies are not shown in \cref{fig:coupling_ftridyn_xolotl} for clarity of exposition.

In the remainder of this section, we will briefly summarize the role of \FTridyn{} and \Xolotl{} in the framework shown in \cref{fig:coupling_ftridyn_xolotl}, with an emphasis on the relevant uncertain parameters in both.

\subsection{The \FTridyn{} binary collision approximation code}\label{sec:the_xolotl_binary_collision_approximation_code}

An efficient simulation method to compute the penetration depth of, and defect production by, energetic particles in solids is the binary collision approximation (BCA). In BCA, particles are assumed to travel through the material in straight paths, undergoing elastic binary collisions with the atoms of the solid. In every collision, the classical scattering integral between the two colliding particles is solved in order to obtain the scattering angle and the amount of energy loss~\cite{smith1997}. A binding energy $E_s$ of the incoming particle to the material can also be taken into account, by adding it to the computed energy loss. The history of a moving particle of species $k$ is terminated when it has been slowed down to an energy below a predefined cutoff energy $E_{f, k}$ (equal to or lower than the surface binding energy). 

In the Monte Carlo BCA method, the distance to the next collision, as well as the impact parameter of the collision, are chosen from a predefined probability distribution. By simulating the trajectory of many independent particles, one can obtain approximations for the sought-after material properties. This is the basis of the \texttt{TRIDYN} code introduced in~\cite{moller1988}. \texttt{TRIDYN} is based on the Monte Carlo program \texttt{TRIM}~\cite{biersack1980}. \FTridyn{} is a flexible code for simulating atomic-scale ion-surface interactions, and can be used to predict sputtering and implantation profiles in the context of plasma-surface interactions~\cite{drobny2018}. We note that \FTridyn{} improves upon \texttt{TRIDYN} by including a robust, fractal model of the surface and additional output modes for coupling to other plasma codes, such as \Xolotl{}. Only this last capability is of interest in this work, however.

In our numerical results presented in \cref{sec:results_and_discussion}, where we consider helium ions irradiated on a tungsten surface, we will define the surface binding energy of tungsten ($E_s$), as well as the cutoff energies of helium and tungsten ($E_{f, \ce{He}}$ and $E_{f, \ce{W}}$) as uncertain parameters. Furthermore, when considering the \PISCES{} problem setting, the IEAD reduces to a single value for the energy $E_\mathrm{in}$ and angle $\alpha_\mathrm{in}$ of the incoming helium flux, which we will also consider as uncertain operators to \FTX{}.

Finally, we note that, since \FTridyn{} is a particle-based Monte Carlo code, it is highly parallelizable. We will exploit this property in our numerical experiments.

\subsection{The \Xolotl{} cluster dynamics code}\label{sec:the_xolotl_cluster_dynamics_code}

\Xolotl{} is a general-purpose cluster dynamics simulator used to predict gas bubble and defect evolution in solids. The code predicts the evolution of a concentration field $C_k(\bsz, t)$ of a species $k$ as a function of space $\bsz$ and time $t$ by solving the spatio-temporal drift-diffusion reaction equations~\cite{faney2014}. We consider four different types of clusters: tungsten self-interstitials ($\ce{I}_l$), lattice vacancies ($\ce{V}_n$), free helium ($\ce{He}_m$), and vacancy-trapped helium ($\ce{He}_m\ce{V}_n$). In these expressions, subscripts indicate the number of atoms in each cluster. Let $\mathscr{K}$ denote the set of all possible species $k$. Assuming symmetry in directions perpendicular to the irradiation flux, the evolution of the concentration $C_k(z, t)$ of each species $k$ in the now one-dimensional geometry is governed by a system of partial differential equations (PDEs) of the form
\begingroup
\allowdisplaybreaks
\begin{align}\label{eq:xolotl_governing_equations}
    \frac{\partial C_k}{\partial t}(z, t) &= D_k \frac{\partial^2 C_k}{\partial z^2}(z, t) \\
    &+ \textrm{Generation}_k(z) \\
    &+ \textrm{Reaction}_k(C_k(z, t) \textrm{ for each } k \in \mathscr{K}) \\
    &+ \textrm{Dissociation}_k(C_k(z, t) \textrm{ for each } k \in \mathscr{K})
\end{align}
\endgroup
for $k \in \mathscr{K}$~\cite{faney2014}. In this expression, the first term on the right-hand side represents the diffusion of the species due to the concentration gradient, with $D_k$ the diffusion coefficient of species $k$. The second term is a generation or production term, representing the change in concentration due to irradiation. The third term represents the change in concentration due to interactions between different types of clusters, and the fourth term represents the change in concentration due to the splitting (thermal emission) of clusters into smaller ones.

Self-interstitials ($\ce{I}_l$), single vacancies ($\ce{V}_1$) and small helium clusters ($\ce{He}_1$ to $\ce{He}_7$) are modeled as mobile species, and diffuse isotropically in the bulk subsurface. Vacancy-trapped helium species ($\ce{He}_m\ce{V}_n$) represent the gas bubbles that grow in the subsurface of the material. The self-interstitials diffuse to the surface and become so-called \textit{adatoms} that grow the tungsten surface.

\Cref{eq:xolotl_governing_equations} is discretized using finite differences in space, on a variable-size grid with increasing resolution near the surface. An implicit Runge--Kutta time integration scheme has been used in \cite{faney2014} due to the stiffness of the diffusion and reaction operators, and will be used here. For fusion-relevant conditions, the number of species $|\mathscr{K}|$, and hence the number of cluster dynamics equations in \eqref{eq:xolotl_governing_equations}, can grow very large, as larger helium-vacancy clusters are required to resolve bubble growth. As noted in~\cite{faney2014}, the nonlinear system in \cref{eq:xolotl_governing_equations} can be split in two parts. The first part describes the movement of the limited number of small, mobile clusters. In this case, the system of equations is typically very small, and it does not grow with the number of species considered. The second part describes the vast number of larger, immobile clusters. In this case, the system of equations reduces to a system of ordinary differential equations (ODEs) that can be trivially parallelized over the different grid points in the domain. For more details on the governing equations underlying \Xolotl{}, we refer to~\cite{maroudas2016, blondel2017}. In our numerical experiments in \cref{sec:results_and_discussion}, the governing equations of \Xolotl{} are solved using \texttt{PETSc}, a scalable partial differential equation solver~\cite{petsc,petsc-web-page}.

The diffusion coefficient $D_k$ of a species $k$ is modeled using an Arrhenius relation, 
\begin{equation}
    D_{k} = D_{k, 0} \exp\left(-\frac{E_k}{k_BT}\right),
\end{equation}
where $E_k$ is the migration energy of the species, $D_{k, 0}$ is the pre-exponential factor, $k_B$ is the Boltzmann constant ($k_B \approx \SI{1.38e-23}{\joule\per\kelvin}$) and $T$ is temperature. In~\cite{faney2014}, the pre-exponential factor and migration energy for each cluster are estimated from large-scale MD simulations. We define the migration energies of all mobile species (i.e., $\ce{He}_1$ to $\ce{He}_7$, and $\ce{V}_1$) as uncertain. On the other hand, we ignore the uncertainties in the migration energies of the self-interstitials $\ce{I}_l$, which are also mobile, because they diffuse much faster compared to the other clusters and do not interact with helium in this model.

The reaction and dissociation terms in the right-hand side of~\eqref{eq:xolotl_governing_equations} are obtained from chemical reaction theory. For two reacting species $k_1$ and $k_2$, the reaction rate $\kappa_{1, 2}$ can be calculated using diffusion-limited reaction theory~\cite{faney2014}. In particular, we have that
\begin{equation}\label{eq:reaction_rates}
    \kappa_{1, 2} = 4 \pi (r_{k_1} + r_{k_2}) (D_{k_1} + D_{k_2})
\end{equation}
where $D_{k_1}$ and $D_{k_2}$ are the diffusion coefficients, and $r_{k_1}$ and $r_{k_2}$ are the \emph{capture radii} of species $k_1$ and $k_2$. For pure helium clusters, the capture radius can be computed from that of the $\ce{He}_1$ cluster, while for interstitial clusters, the capture radius is computed from the capture radii of $\ce{He}_1$ and $\ce{V}_1$, see~\cite[equations (9) and (10)]{faney2014}. The capture radius of $\ce{V}_1$ is multiplied by an additional bias factor in order to reflect that interstitial clusters have a larger surrounding strain field. Both the radius of the $\ce{He}_1$ cluster ($r_{\ce{He}_1}$) and the bias factor ($b_{\ce{I}_l}$) are defined as uncertain parameters, because they directly influence the reaction rates $\kappa_{1, 2}$ in~\eqref{eq:reaction_rates}.

A simplified helium bubble bursting model is included to take into account the gas release mechanisms from near-surface, over-pressurized bubbles. To model the bursting process of these over-pressurized bubbles as reported in \cite{sefta2013, hammond2018}, \Xolotl{} uses a stochastic process. At the end of each time step of the finite difference scheme, the total quantity of mobile and trapped helium atoms $n_\mathrm{He}$ is computed at each grid point. Next, the radius of a helium bubble is computed as
\begin{equation}\label{eq:bubble_radius}
    r_\mathrm{bubble} = a_0 \left( \frac{\sqrt{3}}{4} + \sqrt[3]{\frac{3 n_\mathrm{He}}{32\pi}} - \sqrt[3]{\frac{3}{8\pi}} \right),
\end{equation}
where $a_0$ is a lattice parameter, and assuming perfectly spherical bubbles~\cite{faney2014}. When the distance between the bubble boundary and the surface is non-positive, we assume the bubble to have burst. Otherwise, a bubble may burst with probability
\begin{equation}\label{eq:bubble_bursting_probability}
    p_\mathrm{burst} \propto \left(\frac{r_\mathrm{bubble}}{r}\right) \cdot \gamma \cdot \min\left(1, \exp\left(-\frac{r - \tau_r}{2\tau_r}\right)\right),
\end{equation}
where $r$ is the distance of the center of the bubble to the surface, $\tau_r$ is a depth parameter that prevents bursting from happening deeper in the material, and $\gamma$ is a tuning parameter that ensures the bursting frequency matches the results predicted by a large-scale MD simulation~\cite{blondel2017}. In the present study, we will define $a_0$ as uncertain, while $\tau_r$ and $\gamma$ are defined as deterministic. The bubble bursts are assumed to be pinhole events, in the sense that, after a bubble bursting event occurs, the concentration of vacancy-trapped helium ($\ce{He}_m\ce{V}_n$) is transferred to the concentration of vacancies ($\ce{V}_n$) as if the helium dissipates instantaneously to the surface. Finally, we also include the initial vacancy concentration $C_{\ce{V}_1}$ as an uncertain parameter. The initial vacancy concentration is used as a proxy for material impurities, as commercial tungsten is not perfect.

For computational tractability, the maximum cluster size of the species is limited. We express this limitation in terms of a \emph{network size}, equal to the allowed maximum number of vacancies $n$ in a cluster, imposed for both free vacancies ($\ce{V}_n$) and vacancy-trapped helium ($\ce{He}_m\ce{V}_n$). Finally, to avoid an exponential growth in the number of clusters to be tracked, \Xolotl{} uses a grouping scheme in which clusters of similar size are grouped together into \emph{super-clusters} with averaged properties, and the governing equations are solved for these clusters instead. See also~\cite{kohnert2016} for the first-order moment scheme on which this grouping is based.

\section{Global Sensitivity Analysis Framework}\label{sec:global_sensitivity_analysis_framework}

In this section, we outline our framework for global sensitivity analysis using polynomial chaos expansion (PCE) surrogate models. As a well-established tool for uncertainty representation of and propagation through computational models, PCEs have gained widespread popularity over the past three decades, see, e.g.,~\cite{ghanem1991,lemaitre2010}. In this work, we resort to PCE surrogates because they offer two attractive features. First, they can be used as a surrogate to replace a computationally expensive model in studies that require intensive sampling. Second, owing to the orthogonality of the polynomials, PCE surrogates allow closed-form expressions for variance-based sensitivity indices. We exploit both features in our numerical results presented in~\cref{sec:results_and_discussion}.

\subsection{Polynomial chaos expansion surrogate models}\label{sec:polynomial_chaos_expansion_surrogate_models}

Suppose that, for a given physical model $f$, the relation between the model output $y$ and the $d$-dimensional input parameters $\bsx = (x_1, x_2, \ldots, x_d)$ can be written as
\begin{equation}\label{eq:model}
    y = f(\bsx).
\end{equation}
In what follows, we will refer to $y$ as the \emph{quantity of interest}. Furthermore, in the context of this paper, we assume that the only information available about any $x_j$ is its lower bound $a_j$ and upper bound $b_j$, $j=1, 2, \ldots, d$. In this case, each $x_j$ can be rescaled to a random variable variable $\xi_j \in [-1, 1]$, where 
\begin{equation}\label{eq:mapping}
    \xi_j = 2 \left( \frac{x_j - a_j}{b_j - a_j} \right) - 1, \quad j = 1, 2, \ldots, d.
\end{equation}
A PCE surrogate for $y$ can be written as
\begin{equation}\label{eq:pce_infinite}
    \tilde{y} = \sum_{\bsu \in \calI_d} c_{\bsu} \Phi_\bsu(\bsxi),
\end{equation}
where $\bsu = (u_1, u_2, \ldots, u_d) \in \calI_d$ is a multi-index of length $d$, $\calI_d \subset \N_0^d$ is an appropriately-chosen set of multi-indices, $\Phi_\bsu$ is a multivariate orthogonal polynomial expressed in terms of the rescaled inputs $\bsxi = (\xi_1, \xi_2, \ldots, \xi_d)$, and the $c_\bsu$ are deterministic coefficients~\cite{ghanem2003, smith2013}.

In our setting, the multivariate orthogonal polynomials $\Phi_\bsu$ correspond to a product of (normalized) univariate Legendre polynomials, i.e.,
\begin{equation}
    \Phi_\bsu(\bsxi) \coloneqq \prod_{j=1}^d \phi_{u_j}(\xi_j),
\end{equation}
where $\phi_{u_j}$ is the (normalized) one-dimensional Legendre polynomial of degree $u_j$, $j=1, 2, \ldots, d$, see, e.g.,~\cite{sargsyan2016}. The first four normalized univariate Legendre polynomials $\phi_m(\xi)$ are given by
\begin{align}
    \phi_0(\xi) &= z_0,\\
    \phi_1(\xi) &= z_1 \xi,\\
    \phi_2(\xi) &= \frac{z_2}{2}(3\xi^2 - 1), \text{ and } \\
    \phi_3(\xi) &= \frac{z_3}{2}(5\xi^3 - 3\xi),
\end{align}
where $z_m = \sqrt{2m+1}$ is the normalization constant. By convention, the \emph{order} $|u|$ of the multivariate polynomial $\Phi_\bsu$ is given as the sum of the orders of all univariate polynomials in the expansion, i.e., $|u| \coloneqq u_1 + u_2 + \ldots + u_d$.

An essential part of the PCE construction process is the choice of the index set $\calI_d$. Suppose we assign an index $k$ to each multi-index $\bsu^{(k)} = (u_1^{(k)}, u_2^{(k)}, \ldots, u_d^{(k)})$ in the index set $\calI_d$ and let $K = |\calI_d|$ denote the total number of terms in the expansion. In that case, we can write the PCE as 
\begin{equation}\label{eq:pce_finite}
    \tilde{y} = \sum_{k = 1}^K c_{k} \Phi_k(\bsxi),
\end{equation}
where the $c_{k}$ are the coefficients and $\Phi_k$ is the multivariate polynomial associated with multi-index $\bsu^{(k)}$. In a preprocessing step, one may then identify important parameters or important interaction terms to decide on suitable truncation rules for constructing the index set $\calI_d$. The vector of coefficients $\bsc = (c_k)_{k=1}^K$ can subsequently be determined from a set of input-output pairs $\mathscr{D} = \{(\bsx^{(n)}, y^{(n)})\}_{n=1}^N$ by solving a least-squares regression problem of the form
\begin{equation}
    \bsc^{\mathrm{LS}} = \argmin_\bsc \sum_{n=1}^N \left( y^{(n)} - \sum_{k=1}^K c_k \Phi_k(\bsxi^{(n)}) \right)^2, \label{eq:least_squares}
\end{equation}
where $\bsxi^{(n)}$ corresponds to the rescaled input parameters $\bsx^{(n)}$ defined in~\eqref{eq:mapping}.

In a high-dimensional setting, i.e., when he number of input parameters $d$ is large, the number of basis terms $K$ retained using such a predefined truncation rule may lead to an underdetermined regression problem, especially when only a small number of measurements $N$ are available. This naturally leads to overfitted model approximations, meaning that the model is too complex and has too many degrees of freedom for the data at hand, and eventually to inaccurate sensitivity information extracted from these models. While there are dimension truncation schemes, such as the Smolyak construction~\cite{smolyak1963,conrad2013}, hyperbolic cross construction~\cite{blatman2011} or dimension-adaptive construction~\cite{gerstner2003} that somewhat overcome the dramatic growth of the number of basis terms and work well for low to moderate dimensional problems, these methods are impractical for the 17-dimensional \FTX{} setting at hand. In \cref{sec:bayeseian_compressive_sensing}, we discuss an adaptive approach for constructing the sparse, high-dimensional multi-index set $\calI_d$ based on Compressive Sensing (CS), see~\cite{candes2006,donoho2006}, that overcomes the aforementioned overfitting problem.

\begin{table*}[t]
    \centering
    \begin{tabular}{lll S[table-format=3.3] S[table-format=3.3] S[table-format=3.3] l} \toprule
        & parameter name & symbol & {nominal value} & {lower bound} & {upper bound} & unit \\ \midrule
        \multirow{5}{*}{\rotatebox{90}{\FTridyn{}}}
& surface binding energy W & $E_s$ & 8.78 & 8.68 & 12 & \si{\eV} \\
& cutoff energy W & $E_{f, \ce{W}}$ & 3.0 & 2.7 & 3.3 & \si{\eV} \\
& cutoff energy He & $E_{f, \ce{He}}$ & 0.10 & 0.09 & 0.11 & \si{\eV} \\
& ion impact energy & $E_\mathrm{in}$ & 250 & 240 & 300 & \si{\eV} \\
& incident angle & $\alpha_\mathrm{in}$ & 0 & -30 & 30 & deg \\
& & & & & &\\
\multirow{12}{*}{\rotatebox{90}{\Xolotl{}}}
& lattice parameter & $a_0$ & 0.317 & 0.316 & 0.318 & \si{\nano\metre} \\
& \ce{He_1} radius & $r_{\ce{He_1}}$ & 0.3 & 0.27 & 0.33 & \si{\nano\metre} \\
& $\ce{I}_l$ bias factor & $b_{\ce{I}_l}$ & 1.15 & 1.035 & 1.265 & -- \\
& $\log_{10}$ initial $\ce{V}_1$ & $C_{\ce{V}_1}$ & -18 & -19 & -17 & \si{\nano\metre\tothe{\text{-}3}} \\
& migration energy $\ce{He}_1$ & $E_{\ce{He}_1}$ & 0.13 & 0.11 & 0.25 & \si{\eV} \\
& migration energy $\ce{He}_2$ & $E_{\ce{He}_2}$ & 0.20 & 0.20 & 0.30 & \si{\eV} \\
& migration energy $\ce{He}_3$ & $E_{\ce{He}_3}$ & 0.25 & 0.25 & 0.40 & \si{\eV} \\
& migration energy $\ce{He}_4$ & $E_{\ce{He}_4}$ & 0.20 & 0.15 & 0.50 & \si{\eV} \\
& migration energy $\ce{He}_5$ & $E_{\ce{He}_5}$ & 0.12 & 0.10 & 0.20 & \si{\eV} \\
& migration energy $\ce{He}_6$ & $E_{\ce{He}_6}$ & 0.30 & 0.30 & 0.45 & \si{\eV} \\
& migration energy $\ce{He}_7$ & $E_{\ce{He}_7}$ & 0.40 & 0.30 & 0.45 & \si{\eV} \\
& migration energy $\ce{V}_1$ & $E_{\ce{V}_1}$ & 1.30 & 1.10 & 1.50 & \si{\eV} \\ \bottomrule
    \end{tabular}
    \caption{Specification of the uncertain parameters used in the \FTX{} setting. The ion impact energy and incident angle are only relevant for the \PISCES {} setting.}
    \label{tab:uncertain_parameters}
\end{table*}

\subsection{Global sensitivity analysis using PCEs}\label{sec:global_sensitivity_analysis_using_PCEs}

Global sensitivity analysis (GSA) methods are used to quantify the sensitivity of the model output $y = f(\bsx)$ to the model input parameters $\bsx$. Variance-based sensitivity analysis using Sobol' indices is one of the most commonly used GSA methods, see~\cite{sobol2001}. The Sobol' indices capture the amount of variance in the model output attributed to a certain (set of) input parameter(s). In this section, we briefly recall the definition of the total effect sensitivity indices, and recall how they can be estimated from the PCE for $y$. For a more in-depth analysis, we refer to~\cite{saltelli2008}.

The \emph{total-effect} Sobol' sensitivity indices are defined as
\begin{equation}\label{eq:total_effect_sensitivity_index}
    S^{\text{total}}_j = 1 - \frac{\V_{\bsx_{\shortminus{}j}}[\E_{x_j}[f(\bsx|\bsx_{\shortminus{}j})]]}{\V[f(\bsx)]},
\end{equation}
where $\V_{\bsx_{\shortminus{}j}}[\;\cdot\;]$ is the variance with respect to all parameters $\bsx$ except the $j$th parameter $x_j$, and $\E_{x_j}[\;\cdot\;]$ is the expected value with respect to the $j$th parameter $x_j$. The total-effect sensitivity index is a measure of the fraction of the total variance of the model output that can be attributed to the $j$th parameter, including its interaction with other input variables, see~\cite{sobol2001}.

The total-effect Sobol' sensitivity indices allow for a natural way to order the parameters $x_j$, $j=1, 2, \ldots, d$, according to their relative contribution to the total output variance. We will exploit this property in \cref{sec:results_and_discussion} to identify a subset of important parameters in \FTX{}.

The Sobol' sensitivities are usually computed using a random sampling approximation for~\eqref{eq:total_effect_sensitivity_index} written in integral form -- the so-called \emph{pick-and-freeze} approach, see, e.g.,~\cite{jansen1999, saltelli2002, saltelli2008}.  However, if the model $f(\bsx)$ is computationally expensive to evaluate, this approach may be computationally infeasible, since typically many model evaluations will be required to achieve reasonable accuracy. Alternatively, the values for the sensitivity indices can be computed \emph{analytically} from the coefficients of a PCE surrogate, exploiting the orthogonality of the PCE basis functions. Thus, instead of directly computing the sensitivity indices in~\eqref{eq:total_effect_sensitivity_index} from the statistics derived from the random samples, we construct a PCE surrogate for the model fitted to these same samples, and extract the total effect indices as a post-processing step from the PCE. The advantage of the PCE approach is that it exploits a presumed smoothness in the quantity of interest. This smoothness assumption improves the precision of the estimated sensitivity indices extracted from the surrogate, but may introduce an additional bias due to the truncation of the polynomial order in the PCE, see, e.g.,~\cite{sudret2008, crestaux2009}.

In particular, the total-effect Sobol' sensitivity indices defined in \cref{eq:total_effect_sensitivity_index} can be obtained as
\begin{align}
    S^{\text{total}}_j &\approx \frac{\sum_{\bsu \in \calI_j^{\mathrm{total}}} c_\bsu^2}{\sum_{\bsu \in \calI_d \setminus \{\bszero\}} c_\bsu^2} \label{eq:total_effect_sensitivity_index_with_pce}
\end{align}
with $\calI_j^{\mathrm{total}} = \{\bsu \in \calI_d : u_j > 0 \}$. Hence, having constructed a PCE surrogate, the sought-after sensitivity indices can be computed using the analytical expression in~\eqref{eq:total_effect_sensitivity_index_with_pce}.

\section{Results and discussion}\label{sec:results_and_discussion}

In this section, we describe the results of our GSA study of \FTX{}. First, in \cref{sec:problem_setup}, we provide additional details on the computational and problem setup. Next, in \cref{sec:gsa_study_for_the_iter_he_setting,sec:gsa_study_for_the_pisces_setting}, we present the sensitivity analysis results for the \ITERHe{} and \PISCES{} settings, respectively.

\subsection{Problem setup}\label{sec:problem_setup}

\FTridyn{} and \Xolotl{} are coupled through the integrated plasma simulation (IPS) framework developed in~\cite{elwasif2010}. IPS is a high-performance computing framework providing resource and data management tools for loosely-coupled simulation components. \FTX{} is connected to IPS with the help of a Python wrapper, which controls the time-stepping behavior of the coupled code, running \FTridyn{} and \Xolotl{} sequentially, providing the necessary file-based coupling between them~\cite{ips-wrappers}. The wrapper also deals with code restarts, and with grid extension for \Xolotl{} when needed. For convenience, we also developed \texttt{pyFTX}, a Python interface to the IPS wrapper, that allows us to control multiple \FTX{} simulations at once, including possible checkpoint/restarts.

We construct PCE surrogates and perform GSA using the uncertainty quantification (UQ) toolkit (\texttt{UQTk})~\cite{debusschere2004, debusschere2017}.  We also implemented the adaptive sparse basis growth procedure outlined in \Cref{alg:wibcs} in \cref{sec:bayeseian_compressive_sensing}, relying on the BCS procedure implemented in \texttt{UQTk}.

As stated in \cref{sec:global_sensitivity_analysis_using_PCEs}, the uncertain input parameters are assumed to follow a uniform distribution between a lower and upper bound given in \cref{tab:uncertain_parameters}. It should be noted that we construct the PCE surrogate models in the transformed space $[-1, 1]^d$ for convenience, and map them back to the original parameter space for the sensitivity analysis, following \cref{eq:mapping}.

In each problem setting, we construct a training data set $\mathscr{D}_\textrm{train} = \{(\bsx_\textrm{train}^{(n)}, y_\textrm{train}^{(n)})\}_{n=1}^{N_\textrm{train}}$ with $N_\textrm{train}$ training samples and a test data set $\mathscr{D}_\textrm{test} = \{(\bsx_\textrm{test}^{(n)}, y_\textrm{test}^{(n)})\}_{n=1}^{N_\textrm{test}}$ with $N_\textrm{test}$ test samples. Inspired by~\cite{bohn2018}, we choose the model inputs in the training data set according to a sparse Clenshaw--Curtis quadrature rule, because this may lead to faster convergence of the training error with respect to the number of training samples. However, because of the noise in the outputs, we don't use the associated quadrature rule, but treat these training points as random samples, arguing that this does not necessarily lead to a larger training error. The number of training and test samples in each problem are reported in \cref{tab:problem_setup}. To assess the accuracy of our PCE surrogate models, we compute the training error $e_\textrm{train}$ as
\begin{align}\label{eq:training_error}
    e_\textrm{train} &= \sqrt{\frac{\sum_{n=1}^{N_\textrm{train}} \left(\sum_{k = 1}^K c_{k} \Phi_k(\bsxi_\textrm{train}^{(n)}) - y_\textrm{train}^{(n)}\right)^2}{\sum_{n=1}^{N_\textrm{train}} (y_\textrm{train}^{(n)})^2}}
\end{align}
where $\bsxi_\textrm{train}^{(n)}$ corresponds to the rescaled training input parameters according to~\eqref{eq:mapping}. The test error $e_\textrm{test}$ is computed analogously.

\begin{table}[t]
    \centering
    \small
    \begin{tabular}{l l S[table-format=2] S[table-format=3] S[table-format=3] c} \toprule
        {setting} & {code} & {$d$} & {$N_\textrm{train}$} & {$N_\textrm{test}$} & {cost} \\ \midrule
        \multirow{3}{*}{\ITERHe} & \FTridyn{} & 3 & 441 & 50 & \SI{1}{\hour} \\
        & \Xolotl{} & 12 & 313 & 50 & \SI{12}{\hour} \\
        & \FTX{} & 15 & 481 & 50 & \SI{47}{\hour} \\ \midrule
        \multirow{2}{*}{\PISCES} & \FTX{} & 17 & 613 & 100 & \SI{95}{\hour} \\
        & \FTX{} & 5 & 801 & 100 & \SI{95}{\hour} \\ \bottomrule
    \end{tabular}
    \caption{Number of uncertain parameters $d$, number of training and test samples $N_\textrm{train}$ and  $N_\textrm{test}$, and average computational cost per sample for each problem setting.}
    \label{tab:problem_setup}
\end{table}

We will analyze three outputs of \FTridyn{}: the sputtering yields of tungsten and helium, and the implantation profile. The sputtering yields are scalar output quantities of interest, while the implantation profile changes with the material depth. In the latter case, the depth where the helium implantation reaches its maximum will be the quantity of interest we consider. This depth is obtained from a polynomial fit to a histogram of the particle positions at the simulation end time.

\begin{figure*}[t]
    \centering
	\tikzsetnextfilename{iter_he_f_tridyn}
	\tikzexternalenable
	\setlength{\figurewidth}{5.85cm}
\setlength{\figureheight}{4.75cm}
\begin{tikzpicture}[%
        training/.style={only marks, mark=*, mark size=1pt, RoyalBlue},
        validation/.style={only marks, mark=*, mark size=1pt, Red},
        line/.style={draw, black, dashed, line cap=round},
        label/.style={anchor=north west, font=\scriptsize, xshift=1pt},
        total/.style={ybar, draw=none, fill=Red, bar width=16pt}
    ]
    \begin{groupplot}[%
            group style={%
                group size=3 by 2,
                horizontal sep=1cm,
                vertical sep=3em,
                ylabels at=edge left,
            },
            default axis,
            axis on top,
            legend style={%
                at={(1, 0)},
                anchor=south east,
                /tikz/every odd column/.append style={column sep=1mm}
            },
            clip mode=individual,
            xlabel={model output},
            ylabel={PCE surrogate output},
            title style={%
                align=center,
                font=\small,
            },
            ylabel style={at={(-0.175,0.5)}, anchor=south}
        ]
        \nextgroupplot[%
            xmin=0, xmax=1.5e-3,
            ymin=0, ymax=1.5e-3,
            tick scale binop=\times,
            every x tick scale label/.style={at={(xticklabel cs:1)}, anchor=north},
            every y tick scale label/.style={at={(0,1)}, anchor=south west, yshift=-2pt, xshift=-3pt},
            title={sputtering yield \\ helium [atoms/ion]},
            tick label style={
                /pgf/number format/.cd,
                    fixed,
                    fixed zerofill,
                    precision=1,
                /tikz/.cd
            }
	    ]
        \addplot[training] table[] {dat/iter-he/f-tridyn/parity_sputtering_yield_He_train};
        \addlegendentry{training}
        \addplot[validation] table[] {dat/iter-he/f-tridyn/parity_sputtering_yield_He_val};
        \addlegendentry{test}
        \addplot[line] coordinates {(0, 0) (1.5e-3, 1.5e-3)};
        \node[label, yshift=-1pt] at (rel axis cs:0, 1) {$e_{\mathrm{train}} = 0.018$};
        \node[label, yshift=-11pt] at (rel axis cs:0, 1) {$e_{\mathrm{test}} = 0.025$};
        \node[label, yshift=-21pt] at (rel axis cs:0, 1) {$|u| = 4$};
        \nextgroupplot[%
            xmin=0.18, xmax=0.36,
            ymin=0.18, ymax=0.36,
            xtick={0.18, 0.24, 0.30, 0.36},
            ytick={0.18, 0.24, 0.30, 0.36},
            title={sputtering yield \\ tungsten [atoms/ion]},
            tick label style={
                /pgf/number format/.cd,
                    fixed,
                    fixed zerofill,
                    precision=2,
                /tikz/.cd
            },
	    ]
        \addplot[training] table[] {dat/iter-he/f-tridyn/parity_sputtering_yield_W_train};
        \addplot[validation] table[] {dat/iter-he/f-tridyn/parity_sputtering_yield_W_val};
        \addplot[line] coordinates {(0.18, 0.18) (0.36, 0.36)};
        \node[label, yshift=-1pt] at (rel axis cs:0, 1) {$e_{\mathrm{train}} = 0.001$};
        \node[label, yshift=-11pt] at (rel axis cs:0, 1) {$e_{\mathrm{test}} = 0.001$};
        \node[label, yshift=-21pt] at (rel axis cs:0, 1) {$|u| = 3$};
        \nextgroupplot[%
            xmin=2.7, xmax=3.6,
            ymin=2.7, ymax=3.6,
            xtick={2.7, 3, 3.3, 3.6},
            ytick={2.7, 3, 3.3, 3.6},
            title={depth where helium \\ implantation is maximal [\si{\nano\metre}]},
            tick label style={
                /pgf/number format/.cd,
                    fixed,
                    fixed zerofill,
                    precision=1,
                /tikz/.cd
            }
	    ]
        \addplot[training] table[] {dat/iter-he/f-tridyn/parity_implantation_profile_train};
        \addplot[validation] table[] {dat/iter-he/f-tridyn/parity_implantation_profile_val};
        \addplot[line] coordinates {(2.7, 2.7) (3.6, 3.6)};
        \node[label, yshift=-1pt] at (rel axis cs:0, 1) {$e_{\mathrm{train}} = 0.009$};
        \node[label, yshift=-11pt] at (rel axis cs:0, 1) {$e_{\mathrm{test}} = 0.033$};
        \node[label, yshift=-21pt] at (rel axis cs:0, 1) {$|u| = 8$};
        \nextgroupplot[%
            legend style={%
                at={(1, 1)},
                anchor=north east,
                /tikz/every odd column/.append style={column sep=1mm}
            },
            xmin=-0.5, xmax=2.5,
            ymin=1e-7, ymax=1e1,
            xtick={0, 1, 2},
            xticklabels={$E_s$, $E_{f, \ce{W}}$, $E_{f, \ce{He}}$},
            ytick={1e-7, 1e-5, 1e-3, 1e-1, 1e1},
            ylabel={total sensitivity index},
            ymode=log,
            log origin=infty,
            xlabel={}
	    ]
        \addplot[total] table[x index=0, y index=1] {dat/iter-he/f-tridyn/tot_sensitivity_sputtering_yield_He};
        \nextgroupplot[%
            ymin=1e-10, ymax=1e2,
            ytick={1e-10, 1e-7, 1e-4, 1e-1, 1e2},
            legend style={%
                at={(1, 1)},
                anchor=north east,
                /tikz/every odd column/.append style={column sep=1mm}
            },
            xmin=-0.5, xmax=2.5,
            ymin=1e-10, ymax=1e2,
            ytick={1e-10, 1e-7, 1e-4, 1e-1, 1e2},
            xtick={0, 1, 2},
            xticklabels={$E_s$, $E_{f, \ce{W}}$, $E_{f, \ce{He}}$},
            ymode=log,
            log origin=infty,
            xlabel={}
	    ]
        \addplot[total] table[x index=0, y index=1] {dat/iter-he/f-tridyn/tot_sensitivity_sputtering_yield_W};
        \nextgroupplot[%
            legend style={%
                at={(1, 1)},
                anchor=north east,
                /tikz/every odd column/.append style={column sep=1mm}
            },
            xmin=-0.5, xmax=2.5,
            ymin=1e-7, ymax=1e1,
            xtick={0, 1, 2},
            xticklabels={$E_s$, $E_{f, \ce{W}}$, $E_{f, \ce{He}}$},
            ytick={1e-7, 1e-5, 1e-3, 1e-1, 1e1},
            ymode=log,
            log origin=infty,
            xlabel={}
	    ]
        \addplot[total] table[x index=0, y index=1] {dat/iter-he/f-tridyn/tot_sensitivity_implantation_profile};
    \end{groupplot}
\end{tikzpicture}
	\tikzexternaldisable

    \caption{Comparison of the predicted outputs from the surrogate and the actual model outputs (\emph{top}) and total sensitivity indices (\emph{bottom}) for sputtering yields of helium (\emph{left}) and tungsten (\emph{middle}), and the depth where the helium implantation in the material is maximal (\emph{right}) for \FTridyn{} in the \ITERHe{} setting. We indicate the relative training error $e_\mathrm{train}$, the relative test error $e_\mathrm{test}$, and the order of the PCE $|u|$.}
    \label{fig:iter_he_f_tridyn}
\end{figure*}
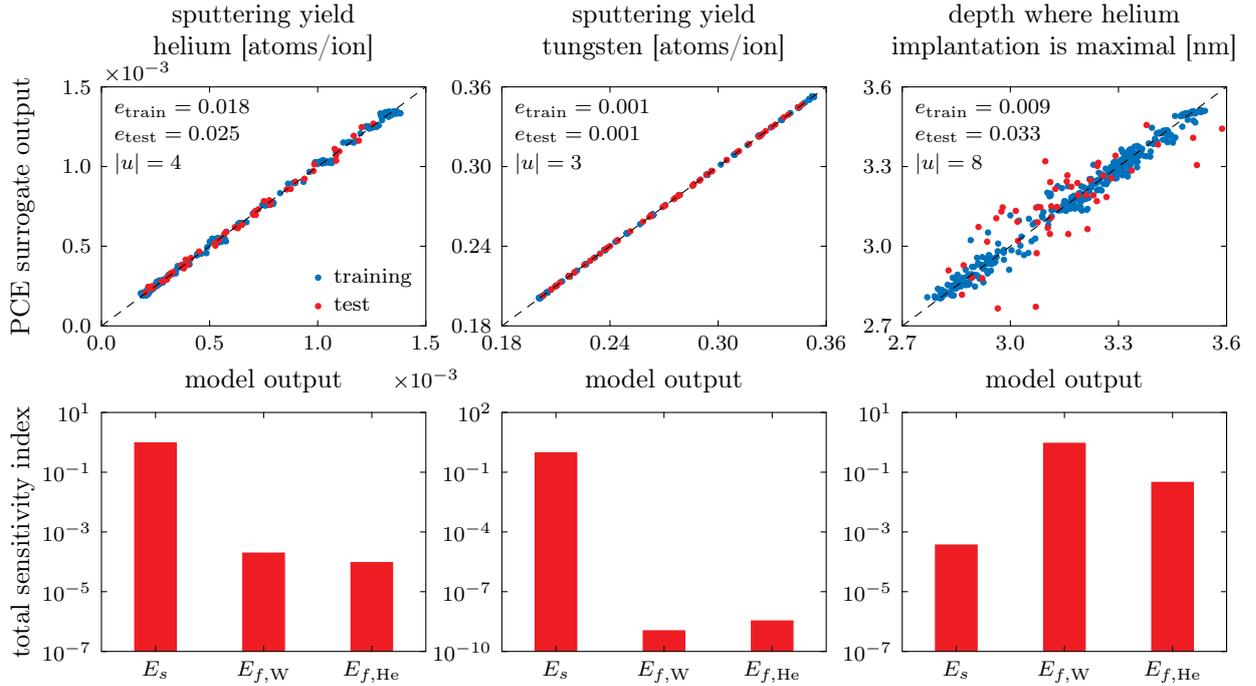

Similarly, we will analyze two outputs of \Xolotl{}: the surface growth and the helium retention. Both output quantities are time-dependent. For the surface growth, we construct a total of $M_t = 25$ surrogates at time stamps $t_m = m/M_t$ for $m=1, 2, \ldots, M_t$. Note that the initial height of the surface is assumed to be constant for all samples, so no surrogate needs to be constructed at $t = \SI{0}{\second}$. The helium retention is expressed as a percentage with respect to the implanted helium flux. Because it contains the effect of the individual bubble bursting events, we first smooth the output using a low-pass filter with a window size of $\delta t = \SI{0.1}{\second}$. Next, we construct $M_t = 25$ surrogates for the smoothed helium retention at time stamps $t_m = \delta t + (m - 1)(1 - \delta t)/(M_t - 1)$ for $m=1, 2, \ldots, M_t$.

In the \ITERHe{} setting, we select an incoming helium flux of \SI{3.49e20}{\per\metre\squared\per\second} incident on a (100)-oriented tungsten surface at \SI{343}{\kelvin}, based on our previous integrated simulations of the early stage of planned ITER operation with a helium plasma~\cite{blondel2017a}. These conditions correspond to the location with peak plasma temperature predicted by simulation in~\cite{lasa2020}. We assume no heat flux is applied to the surface in this case, because the change in surface temperature for this location, once the steady state is reached, is less than \SI{2}{\kelvin}, which is negligible for our study. Moreover, the inclusion of surface heat flux would negatively impact the computational cost, see~\cite{lasa2020}. Furthermore, we use a material with a thickness of \SI{6}{\milli\metre} to mimic the distance between the PFC surface and the cooling channel of a divertor tile~\cite{pitts2017}. We obtain the IEAD, required by \FTridyn{}, from a simulation with \texttt{hPIC}~\cite{khaziev2015}.

In the \PISCES{} setting, we select an incoming helium flux of \SI{540e20}{\per\metre\squared\per\second} incident on a (111)-oriented tungsten surface at \SI{1093}{\kelvin} with a thickness of \SI{5}{\micro\metre}, and with a biasing voltage of \SI{250}{\volt}. A value of \SI{5}{\micro\metre} was selected as it is much larger than the helium diffusion distance for this timescale. For comparison, the radius of the largest bubble we consider in this case is \SI{0.96}{\nano\metre}. In the \PISCES{} setting, the IEAD reduces to a single value for the ion impact energy and angle, and both energy and angle are considered as additional uncertain operating conditions, see \cref{tab:uncertain_parameters}.

In both the \ITERHe{} and \PISCES{} settings, we simulate an exposure time of \SI{1}{\second}. \FTridyn{} is run for a total of $10^5$ particles. We solve the one-dimensional governing equations in \Xolotl{} on a grid with 367 cells, and assuming a network size of at most 250 vacancies ($\ce{V}_n$), which amounts to 13,113 degrees of freedom for each cell.

\subsection{GSA study for the \ITERHe{} setting}\label{sec:gsa_study_for_the_iter_he_setting}

We now report the results of our GSA study in the \ITERHe{} setting. We first consider \FTridyn{} and \Xolotl{} in isolation, before presenting the results for the coupled \FTX{} case.

\subsubsection{Sensitivity analysis of \FTridyn{}}\label{sec:sensitivity_analysis_of_f-tridyn}

In the top row of \cref{fig:iter_he_f_tridyn}, we show a comparison of the predicted \FTridyn{} outputs from the PCE surrogates with the actual model outputs. We also report the corresponding relative training and test errors $e_\textrm{train}$ and $e_\textrm{test}$ computed using~\eqref{eq:training_error}, and the order of the PCE $|u|$. Because the training and test errors are close to each other, we conclude that the PCE surrogates are not overfitted. Furthermore, note that the surrogate models for the sputtering yields are more accurate compared to the surrogate model for the implantation profile. The total order of the PCE ($|u|=8$) for the latter quantity is much larger, indicating a strong nonlinear dependence of the output quantity on the respective the input parameters.

In the bottom row of \cref{fig:iter_he_f_tridyn}, we report the total sensitivity indices for each output. For the sputtering yields, almost all the uncertainty in the output can be explained by the uncertainty in the surface binding energy of tungsten (note the logarithmic scale). For the depth where the helium implantation profile has a maximum, both the surface binding energy of tungsten and the cutoff energy of tungsten appear to be important.

\begin{figure*}
    \centering
	\tikzsetnextfilename{iter_he_f_tridyn_convergence}
	\tikzexternalenable
	\setlength{\figurewidth}{5.85cm}
\setlength{\figureheight}{4.9cm}
\pgfplotstableread[header=false]{dat/iter-he/f-tridyn/convergence_tot_sens_implantation_profile}\data
\begin{tikzpicture}[%
        order1/.style={default marker line, NavyBlue},
        order2/.style={default marker line, BrickRed},
        wibcs/.style={default marker line, ForestGreen}
    ]
    \begin{groupplot}[%
            group style={%
                group size=3 by 1,
                horizontal sep=2em,
                ylabels at=edge left,
                yticklabels at=edge left,
            },
            default axis,
            axis on top,
            legend style={%
                at={(1, 1)},
                anchor=north east,
                /tikz/every odd column/.append style={column sep=1mm}
            },
            clip mode=individual,
            xmin=100, xmax=950,
            ymin=1e-5, ymax=1e0,
            ymode=log,
            xlabel={$N_{\textrm{train}}$},
            ylabel={error in total sensitivity},
            title style={%
                align=center,
                yshift=-5pt,
            },
            xtick={100,300,500,700,900},
        ]
        \nextgroupplot[%
            title={$E_s$}
        ]
        \addplot[order1, default error] table[y error minus index=2, y error plus index=3] {dat/iter-he/f-tridyn/totsens_error_SBV_W_0.dat};
        \addlegendentry{order 1}
        \addplot[order2, default error] table[y error minus index=2, y error plus index=3] {dat/iter-he/f-tridyn/totsens_error_SBV_W_1.dat};
        \addlegendentry{order 2}
        \addplot[wibcs, default error] table[y error minus index=2, y error plus index=3] {dat/iter-he/f-tridyn/totsens_error_SBV_W_2.dat};
        \addlegendentry{BCS}
        \nextgroupplot[%
            title={$E_{f, \ce{W}}$}
        ]
        \addplot[order1, default error] table[y error minus index=2, y error plus index=3] {dat/iter-he/f-tridyn/totsens_error_EF_W_0.dat};
        \addplot[order2, default error] table[y error minus index=2, y error plus index=3] {dat/iter-he/f-tridyn/totsens_error_EF_W_1.dat};
        \addplot[wibcs, default error] table[y error minus index=2, y error plus index=3] {dat/iter-he/f-tridyn/totsens_error_EF_W_2.dat};
        \nextgroupplot[%
            title={$E_{f, \ce{He}}$}
        ]
        \addplot[order1, default error] table[y error minus index=2, y error plus index=3] {dat/iter-he/f-tridyn/totsens_error_EF_He_0.dat};
        \addplot[order2, default error] table[y error minus index=2, y error plus index=3] {dat/iter-he/f-tridyn/totsens_error_EF_He_1.dat};
        \addplot[wibcs, default error] table[y error minus index=2, y error plus index=3] {dat/iter-he/f-tridyn/totsens_error_EF_He_2.dat};
    \end{groupplot}
\end{tikzpicture}
	\tikzexternaldisable

    \caption{Error in the total effect Sobol' sensitivity index as a function of the number of training samples for $E_s$ (\emph{left}), $E_{f, \ce{W}}$ (\emph{middle}) and $E_{f, \ce{He}}$ (\emph{right}) for \FTridyn{} in the \ITERHe{} setting. The error in the total sensitivity index predicted by BCS is much lower compared to the error predicted by a PCE of order 1 or 2 for the same number of training samples.}
    \label{fig:iter_he_f_tridyn_convergence}
\end{figure*}
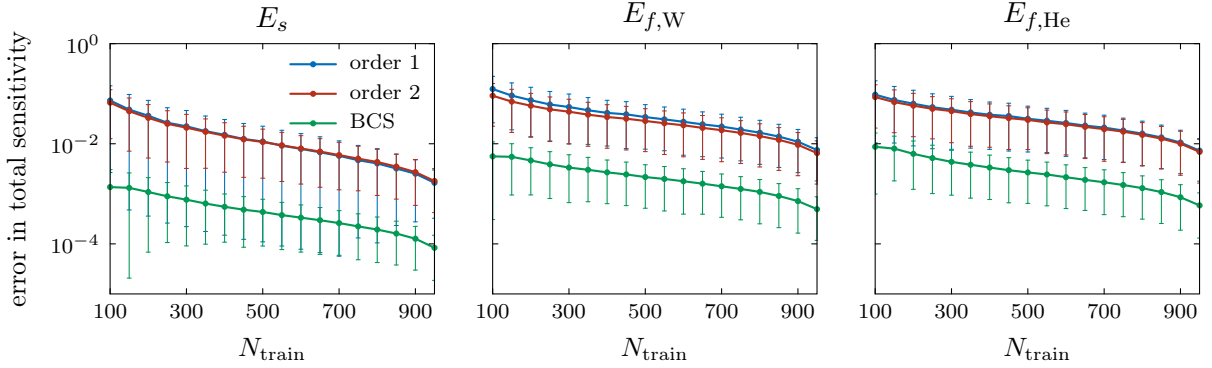

To illustrate the efficacy of our method, we include an additional 500 training samples, and reconstruct the PCE for the maximum helium implantation depth, for an increasing number of training samples $N_{\textrm{train}}$. We consider a PCE surrogate with maximum polynomial order $|u|=1$ and $|u|=2$, as well as BCS with adaptive basis growth. In \cref{fig:iter_he_f_tridyn_convergence}, we check the convergence of the sensitivity indices of these PCEs as a function of $N_{\textrm{train}}$. Note how the error in the total sensitivity index obtained with BCS is consistently lower than the order-1 and order-2 PCE constructions. These results indicate the merits of our higher-order, adaptive basis growth construction in the presence of only a limited amount of training data. 

\subsubsection{Sensitivity analysis of \Xolotl{}}\label{sec:sensitivity_analysis_of_xolotl}

\begin{figure}[t]
    \centering
	\tikzsetnextfilename{iter_he_x}
	\tikzexternalenable
	\setlength{\figurewidth}{6.375cm}
\setlength{\figureheight}{5.25cm}
\pgfplotstableread[header=false]{dat/iter-he/xolotl/parity}{\parity}
\pgfplotstableread[header=false]{dat/iter-he/xolotl/gsa}{\gsa}
\begin{tikzpicture}[%
        default marker/.style={only marks, mark=*, mark size=1pt},
        line/.style={draw, black, dashed, line cap=round},
        label/.style={anchor=north west, font=\scriptsize, xshift=1pt},
        error/.style={default line, black}
    ]
    \begin{groupplot}[%
            group style={%
                group size=2 by 1,
                horizontal sep=3.5cm,
                vertical sep=3em
            },
            default axis,
            axis on top,
            legend style={%
                at={(1.05, 1.04)},
                font={\normalsize},
                anchor=north west,
                /tikz/every odd column/.append style={column sep=1mm}
            },
            clip mode=individual,
            title style={%
                align=center,
                yshift=-5pt,
                font=\small,
            },
            ylabel style={at={(-0.175,0.5)}, anchor=south},
        ]
        \nextgroupplot[%
            xmin=0, xmax=1,
            ymin=0, ymax=1,
            xlabel={model output},
            ylabel={PCE surrogate output},
            colormap/jet,
            colorbar,
            colorbar style={%
                ylabel={time [\si{\second}]},
                ylabel style={at={(2.5,0.5)}, anchor=north},
                ymin=0.1, ymax=1,
                point meta min=0.1, point meta max=1,
                width=1em,
                at={(1.05, 0.5)},
                anchor=west,
                ytick style={draw=none},
                ytick={0.1,0.2,...,1},
                yticklabel style={
                    /pgf/number format/.cd,
                        fixed,
                        fixed zerofill,
                        precision=1
                }
            },
            title={{helium retention} [\%]},
	    ]
        \pgfplotsinvokeforeach{0, 2, ..., 22}{%
            \pgfmathtruncatemacro{\y}{#1 + 1}
            \addplot[default marker, color of colormap={#1/22*1000}] table[x index=#1, y index=\y] \parity;
        }
        \addplot[line] coordinates {(0, 0) (1, 1)};
        \node[label, yshift=-1pt] at (rel axis cs:0, 1) {$e_{\mathrm{train}} = 0.009$};
        \node[label, yshift=-11pt] at (rel axis cs:0, 1) {$e_{\mathrm{test}} = 0.012$};
        \node[label, yshift=-21pt] at (rel axis cs:0, 1) {$|u| = 2$};
        \nextgroupplot[%
            xlabel={time [\si{\second}]},
            ylabel={total sensitivity index},
            xmin=0, xmax=1,
            ymin=1e-6, ymax=1,
            ytick={1e-6, 1e-4, 1e-2, 1e0},
            ymode=log,
	    ]
        \plotsens{5}{Red}{\gsa}{}
        \addlegendentry{$E_{\ce{He}_1}$}
        \plotsens{4}{Dandelion}{\gsa}{}
        \addlegendentry{$r_{\ce{He}_1}$}
        \plotsens{3}{NavyBlue}{\gsa}{}
        \addlegendentry{$E_{\ce{He}_3}$}
        \plotsens{2}{ForestGreen}{\gsa}{}
        \addlegendentry{$E_{\ce{He}_2}$}
        \node[] at (rel axis cs:0.33, 0.7) {$E_{\ce{He}_1}$};
        \node[] at (rel axis cs:0.45, 0.375) {$r_{\ce{He}_1}$};
        \node[] at (rel axis cs:0.7, 0.15) {$E_{\ce{He}_3}$};
        \addplot[error, smooth] table[x index=0, y index=1] {dat/iter-he/xolotl/error.dat};
    \end{groupplot}
\end{tikzpicture}
	\tikzexternaldisable

    \caption{Comparison of the predicted outputs from the surrogate and the actual model outputs for the helium retention (\emph{top}) and total sensitivity indices (\emph{right}) for \Xolotl{} in the \ITERHe{} setting.}
    \label{fig:iter_he_x}
\end{figure}
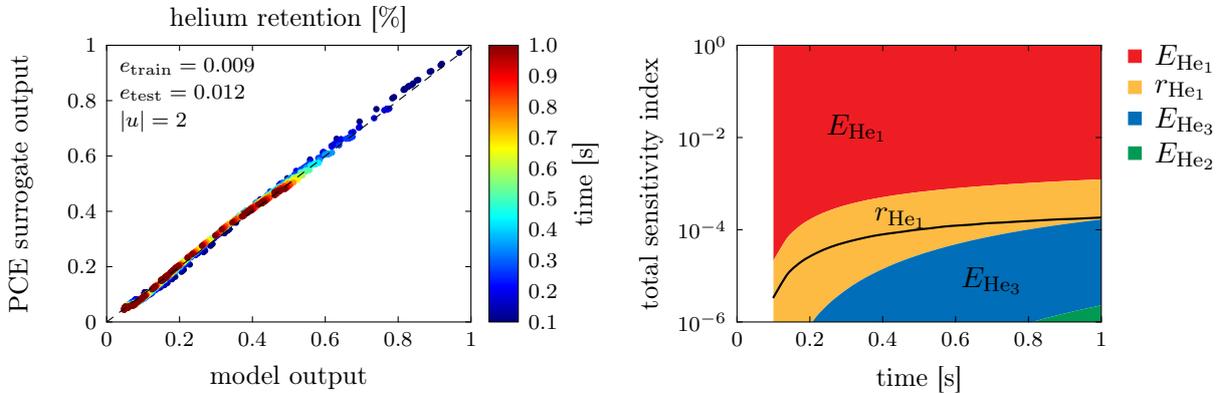

In the top pane of~\cref{fig:iter_he_x} we show a comparison of the helium retention predicted by the PCE surrogate with the actual helium retention predicted by \Xolotl{} as a function of time, as well as the corresponding relative training and test errors, and the PCE order for the surrogate with the largest test error across all time stamps.

In the bottom pane of~\cref{fig:iter_he_x}, we show the total-effect Sobol' sensitivity indices as a function of time, extracted from the PCE surrogates. Almost all uncertainty in the helium retention can be explained by the variation of the migration energy $E_{\ce{He}_1}$ (note the logarithmic $y$-axis), and this seems to be consistent across all time values considered in this case.

The solid line in the bottom pane of~\cref{fig:iter_he_x} is a measure of the reliability of the sensitivity indices in the presence of surrogate error. We computed this threshold by inflating the posterior on the PCE coefficients $c_\bsu$, obtained from BCS, such that the standard deviation of the model predictions with these (now uncertain) surrogates matches the (absolute) validation error. We then plot the maximum value of the standard deviation of the sensitivity predictions as a function of time. In doing so, the surrogate error is effectively reflected in the uncertainty of the sensitivity indices. Values of the sensitivity indices below the threshold are considered to be unreliable due to surrogate error.

As there was no significant change in the surface height within the simulated exposure time of \SI{1}{\second}, a GSA study for the latter is not required.

\subsubsection{Sensitivity analysis of coupled \FTX{}}\label{sec:sensitivity_analysis_of_coupled_ftx}

\begin{figure*}[t]
    \centering
	\tikzsetnextfilename{iter_he_ftx}
	\tikzexternalenable
	\setlength{\figurewidth}{7cm}
\setlength{\figureheight}{5.75cm}
\begin{tikzpicture}[%
        line/.style={draw, black, dashed, line cap=round},
        label/.style={anchor=north west, font=\scriptsize, xshift=1pt},
        error/.style={default line, black},
    ]
    \begin{groupplot}[%
            group style={%
                group size=2 by 2,
                horizontal sep=5em,
                vertical sep=3em,
                xlabels at=edge bottom,
                ylabels at=edge left,
            },
            default axis,
            axis on top,
            legend style={%
                at={(1, 1)},
                anchor=north east,
                /tikz/every odd column/.append style={column sep=1mm}
            },
            clip mode=individual,
            ylabel={PCE surrogate output},
            title style={%
                align=center,
                yshift=-5pt,
                font=\small
            },
            ylabel style={at={(-0.15,0.5)}, anchor=south},
        ]
        \nextgroupplot[%
            xmin=0, xmax=0.4,
            ymin=0, ymax=0.4,
            title={{surface growth} [\si{\nano\metre}]},
            ticklabel style={
                /pgf/number format/.cd,
                    fixed,
                    fixed zerofill,
                    precision=1
            },
            xlabel={model output}
        ]
        \addplot graphics[xmin=0, xmax=0.4, ymin=0, ymax=0.4] {plt/iter-he/ftx/parity_surface_growth.png};
        \addplot[line] coordinates {(0, 0) (0.4, 0.4)};
        \node[label, yshift=-1pt] at (rel axis cs:0, 1) {$e_{\mathrm{train}} = 0.029$};
        \node[label, yshift=-11pt] at (rel axis cs:0, 1) {$e_{\mathrm{test}} = 0.054$};
        \node[label, yshift=-21pt] at (rel axis cs:0, 1) {$|u| = 2$};
        \nextgroupplot[%
            xmin=0, xmax=1,
            ymin=0, ymax=1,
            colormap/jet,
            colorbar,
            colorbar style={%
                ylabel={time [\si{\second}]},
                ylabel style={at={(2.5,0.5)}, anchor=north},
                ymin=0.1, ymax=1,
                point meta min=0.1, point meta max=1,
                width=1em,
                at={(1.05, 0.5)},
                anchor=west,
                ytick style={draw=none},
                ytick={0.1,0.2,...,1},
                yticklabel style={
                    /pgf/number format/.cd,
                        fixed,
                        fixed zerofill,
                        precision=1
                }
            },
            ticklabel style={
                /pgf/number format/.cd,
                    fixed,
                    fixed zerofill,
                    precision=1
            },
            title={{helium retention} [\%]},
            xlabel={model output},
        ]
        \addplot graphics[xmin=0, xmax=1, ymin=0, ymax=1] {plt/iter-he/ftx/parity_he_retention.png};
        \addplot[line] coordinates {(0, 0) (1, 1)};
        \node[label, yshift=-1pt] at (rel axis cs:0, 1) {$e_{\mathrm{train}} = 0.024$};
        \node[label, yshift=-11pt] at (rel axis cs:0, 1) {$e_{\mathrm{test}} = 0.071$};
        \node[label, yshift=-21pt] at (rel axis cs:0, 1) {$|u| = 3$};
        \nextgroupplot[%
            legend style={%
                at={(1.05, 1.04)},
                font={\normalsize},
                anchor=north west,
                /tikz/every odd column/.append style={column sep=1mm}
            },
            clip mode=individual,
            xlabel={time [\si{\second}]},
            ylabel={total sensitivity index},
            xmin=0, xmax=1,
            ymin=1e-2, ymax=1,
            ymode=log,
            ylabel shift=-5pt,
        ]
        \plotsens{15}{Red}{dat/iter-he/ftx/gsa_surface_growth}{}
        \addlegendentry{\strut$E_{\ce{He}_1}$}
        \plotsens{14}{NavyBlue}{dat/iter-he/ftx/gsa_surface_growth}{}
        \addlegendentry{\strut$E_{\ce{He}_2}$}
        \plotsens{13}{Lavender}{dat/iter-he/ftx/gsa_surface_growth}{}
        \addlegendentry{\strut$E_{\ce{He}_4}$}
        \plotsens{12}{Dandelion}{dat/iter-he/ftx/gsa_surface_growth}{}
        \addlegendentry{\strut$r_{\ce{He}_1}$}
        \plotsens{11}{Gray}{dat/iter-he/ftx/gsa_surface_growth}{}
        \addlegendentry{\strut$E_{\ce{He}_5}$}
        \plotsens{10}{ForestGreen}{dat/iter-he/ftx/gsa_surface_growth}{}
        \addlegendentry{\strut$E_{\ce{He}_3}$}
        \node[] at (rel axis cs:0.2, 0.8) {$E_{\ce{He}_1}$};
        \node[] at (rel axis cs:0.15, 0.45) {$E_{\ce{He}_2}$};
        \node[] at (rel axis cs:0.65, 0.425) {$E_{\ce{He}_4}$};
        \node[] at (rel axis cs:0.4, 0.15) {$r_{\ce{He}_1}$};
        \node[] at (rel axis cs:0.86, 0.05) {$E_{\ce{He}_5}$};
        \node[] at (rel axis cs:0.625, 0.05) {$E_{\ce{He}_3}$};
        \addplot[error, smooth] table[x index=0, y index=1] {dat/iter-he/ftx/error-1.dat};
        \nextgroupplot[%
            legend style={%
                at={(1.05, 1.04)},
                font={\normalsize},
                anchor=north west,
                /tikz/every odd column/.append style={column sep=1mm}
            },
            clip mode=individual,
            xlabel={time [\si{\second}]},
            xmin=0, xmax=1,
            ymin=1e-2, ymax=1,
            ymode=log,
            yticklabel=\empty
        ]
        \plotsens{15}{Red}{dat/iter-he/ftx/gsa_he_retention}{}
        \addlegendentry{\strut$E_{\ce{He}_1}$}
        \plotsens{14}{NavyBlue}{dat/iter-he/ftx/gsa_he_retention}{}
        \addlegendentry{\strut$E_{\ce{He}_2}$}
        \plotsens{13}{Lavender}{dat/iter-he/ftx/gsa_he_retention}{}
        \addlegendentry{\strut$E_{\ce{He}_4}$}
        \plotsens{12}{Dandelion}{dat/iter-he/ftx/gsa_he_retention}{}
        \addlegendentry{\strut$r_{\ce{He}_1}$}
        \plotsens{11}{ForestGreen}{dat/iter-he/ftx/gsa_he_retention}{}
        \addlegendentry{\strut$E_{\ce{He}_3}$}
        \plotsens{10}{Gray}{dat/iter-he/ftx/gsa_he_retention}{}
        \addlegendentry{\strut$E_{\ce{He}_5}$}
        \plotsens{9}{Magenta}{dat/iter-he/ftx/gsa_he_retention}{}
        \addlegendentry{\strut$E_{\ce{He}_7}$}
        \node[] at (rel axis cs:0.3, 0.8) {$E_{\ce{He}_1}$};
        \node[] at (rel axis cs:0.325, 0.45) {$E_{\ce{He}_2}$};
        \node[] at (rel axis cs:0.49, 0.325) {$E_{\ce{He}_4}$};
        \node[] at (rel axis cs:0.925, 0.25) {$r_{\ce{He}_1}$};
        \node[] at (rel axis cs:0.45, 0.075) {$E_{\ce{He}_3}$};
        \node[] at (rel axis cs:0.65, 0.125) {$E_{\ce{He}_5}$};
        \addplot[error, smooth] table[x index=0, y index=1] {dat/iter-he/ftx/error-2.dat};
    \end{groupplot}
    \node (title) at ($(group c1r1.center)!0.5!(group c2r1.center)+(0,3cm)$) {\FTX{} \textbf{(15 parameters),} \ITERHe{} \textbf{setting}};
\end{tikzpicture}
	\tikzexternaldisable

    \caption{Comparison of the predicted outputs from the surrogate and the actual model outputs (\emph{top}) and total sensitivity indices as a function of time (\emph{bottom}) for surface growth (\emph{left}) and helium retention (\emph{right}) for \FTX{} in the \ITERHe{} setting.}
    \label{fig:iter_he_ftx}
\end{figure*}
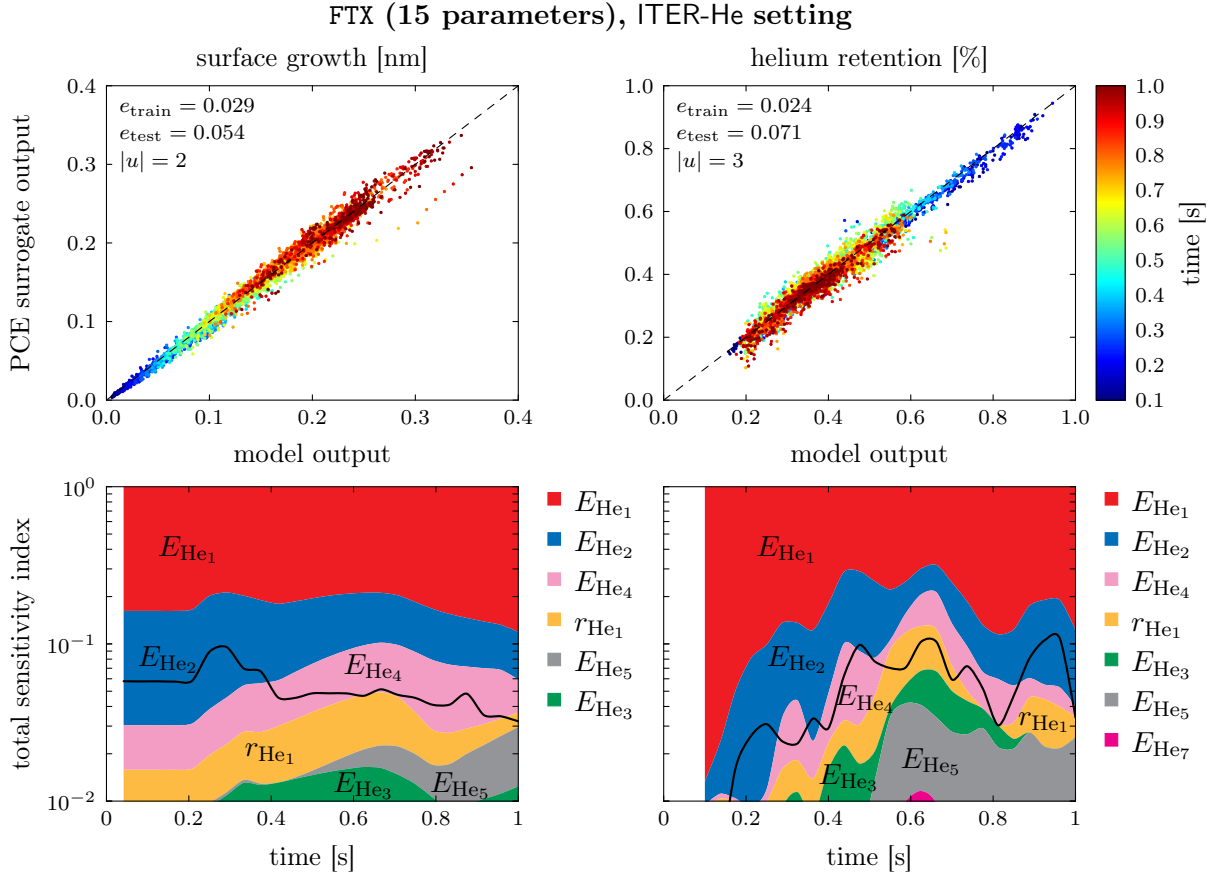

In the top row of~\cref{fig:iter_he_ftx}, we illustrate the accuracy of the surrogate models for both the surface growth and the helium retention predicted by \FTX{}. Different colors indicate different time stamps. The maximum test error is observed at the end time ($\SI{1}{\second}$) for both output quantities of interest.

In the bottom row of~\cref{fig:iter_he_ftx}, we show the total sensitivity indices for the most important parameters as a function of time. The solid line is a measure of the reliability of the sensitivity indices in the presence of surrogate error. Again, the migration energy $E_{\ce{He}_1}$ is the most important parameter, followed by the migration energies $E_{\ce{He}_2}$ and $E_{\ce{He}_4}$, and the impurity radius. This is the case for both the surface growth and the helium retention outputs. The relative importance of $E_{\ce{He}_1}$ seems to decrease as a function of time. Note that this is also the most important parameter for \Xolotl{} in isolation as shown above (\cref{sec:sensitivity_analysis_of_xolotl}). None of the \FTridyn{} parameters seem to have an impact on the surface growth and helium retention predicted by the coupled code in the \ITERHe{} problem setting, indicating that most of the uncertainty in the output is due to variations in the \Xolotl{} parameters. We believe the reason for this is twofold. First, the simulated exposure time of \SI{1}{\second} is not long enough to change the surface height considerably. This value is the best we could achieve given the computational constraints. As a consequence, the change in the value of the sputtering yields, predicted by \FTridyn{} and used as an input to \Xolotl{}, is small. Second, no uncertainties are considered in the IEAD, an input to \FTridyn{}. Together, these two observations explain why the \FTridyn{} parameters have a negligible impact on the helium retention predicted by the \FTX{} in the \ITERHe{} setting.

\subsection{GSA study for the \PISCES{} setting}\label{sec:gsa_study_for_the_pisces_setting}

In this section, we report the results of our GSA study in the \PISCES{} setting, focusing on \FTX{}.

\subsubsection{Sensitivity analysis of coupled \FTX{}}

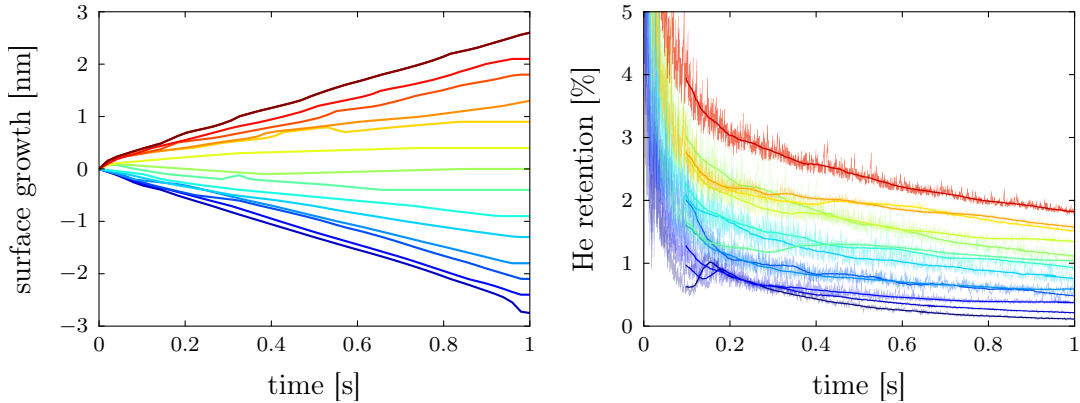
\begin{figure}[t]
    \centering
	\tikzsetnextfilename{pisces_ftx_outputs}
	\tikzexternalenable
	\setlength{\figurewidth}{7.25cm}
\setlength{\figureheight}{5.75cm}
\begin{tikzpicture}[%
    ]
    \begin{groupplot}[%
            group style={%
                group size=2 by 1,
                horizontal sep=1.5cm,
                ylabels at=edge left,
                xlabels at=edge bottom,
                xticklabels at=edge bottom,
                vertical sep=1em,
            },
            default axis,
            axis on top,
            legend style={%
                at={(1, 1)},
                anchor=north east,
                /tikz/every odd column/.append style={column sep=1mm}
            },
            clip mode=individual,
            xlabel={time [\si{\second}]},
            xmin=0, xmax=1,
        ]
        \nextgroupplot[%
            ylabel={surface growth [\si{nm}]},
            ymin=-3, ymax=3,
            ytick={-3, -2, -1, 0, 1, 2, 3},
            colormap/jet
	    ]
        \pgfplotsinvokeforeach{1, 2, ..., 15}{%
            \addplot[default line, color of colormap={#1/15*1000}] table[x index=0, y index=#1] {dat/pisces/ftx/surface_growth.dat};
        }
        \nextgroupplot[%
            ymin=0, ymax=5,
            ytick={0, 1, 2, 3, 4, 5},
            ylabel={He retention [\%]}
	    ]
        \addplot graphics[xmin=0, xmax=1, ymin=0, ymax=5] {plt/pisces/ftx/he_retention_lower_res.png};
    \end{groupplot}
\end{tikzpicture}
	\tikzexternaldisable

    \caption{Select realizations of the surface growth as a function of time (\emph{top}), and helium retention in the tungsten material expressed as a percentage of the implanted helium flux as a function of time (\emph{bottom}), for the \PISCES{} setting. We use a moving average filter to smooth the helium retention before performing GSA.}
    \label{fig:pisces_ftx_outputs}
\end{figure}

In \cref{fig:pisces_ftx_outputs}, we illustrate the surface growth and helium retention as predicted by \FTX{} as a function of time, for select parameter values. Note that the computational burden per sample in the \PISCES{} setting, reported in \cref{tab:problem_setup}, is almost twice that of the \ITERHe{} setting. This results from the higher incoming helium flux, which causes more bursting events with associated smaller time steps.

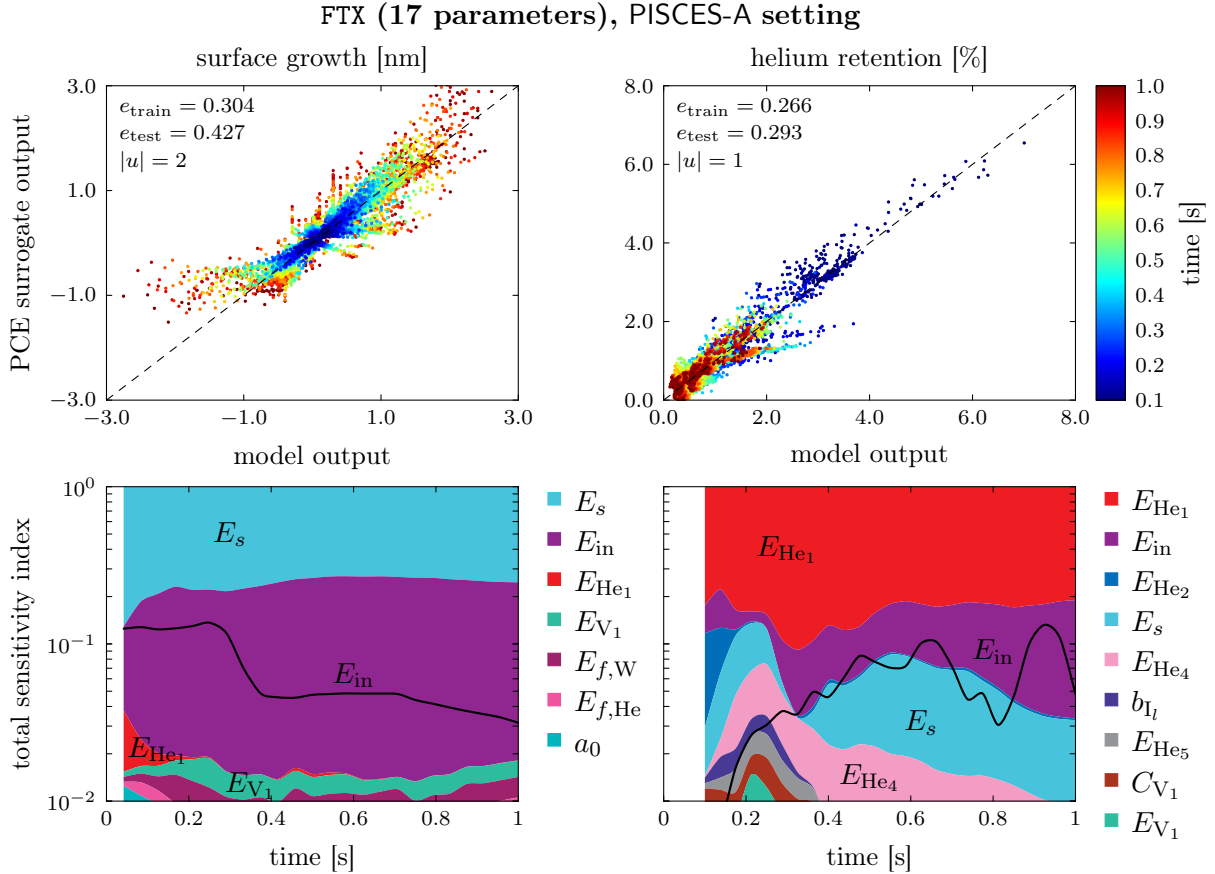
\begin{figure*}[t]
    \centering
	\tikzsetnextfilename{pisces_ftx}
	\tikzexternalenable
	\setlength{\figurewidth}{7cm}
\setlength{\figureheight}{5.75cm}
\begin{tikzpicture}[%
        line/.style={draw, black, dashed, line cap=round},
        label/.style={anchor=north west, font=\scriptsize, xshift=1pt},
        error/.style={default line, black}
    ]
    \begin{groupplot}[%
            group style={%
                group size=2 by 2,
                horizontal sep=5em,
                vertical sep=3em,
                xlabels at=edge bottom,
                ylabels at=edge left,
            },
            default axis,
            axis on top,
            legend style={%
                at={(1, 1)},
                anchor=north east,
                /tikz/every odd column/.append style={column sep=1mm}
            },
            clip mode=individual,
            ylabel={PCE surrogate output},
            title style={%
                align=center,
                yshift=-5pt,
                font=\small
            },
            ylabel style={at={(-0.15,0.5)}, anchor=south},
        ]
        \nextgroupplot[%
            xmin=-3, xmax=3,
            ymin=-3, ymax=3,
            xtick={-3, -1, ..., 3},
            ytick={-3, -1, ..., 3},
            title={{surface growth} [\si{\nano\metre}]},
            ticklabel style={
                /pgf/number format/.cd,
                    fixed,
                    fixed zerofill,
                    precision=1
            },
            xlabel={model output}
        ]
        \addplot graphics[xmin=-3, xmax=3, ymin=-3, ymax=3] {plt/pisces/ftx/parity_surface_growth.png};
        \addplot[line] coordinates {(-3, -3) (3, 3)};
        \node[label, yshift=-1pt] at (rel axis cs:0, 1) {$e_{\mathrm{train}} = 0.304$};
        \node[label, yshift=-11pt] at (rel axis cs:0, 1) {$e_{\mathrm{test}} = 0.427$};
        \node[label, yshift=-21pt] at (rel axis cs:0, 1) {$|u| = 2$};
        \nextgroupplot[%
            xmin=0, xmax=8,
            ymin=0, ymax=8,
            colormap/jet,
            colorbar,
            colorbar style={%
                ylabel={time [\si{\second}]},
                ylabel style={at={(2.5,0.5)}, anchor=north},
                ymin=0.1, ymax=1,
                point meta min=0.1, point meta max=1,
                width=1em,
                at={(1.05, 0.5)},
                anchor=west,
                ytick style={draw=none},
                ytick={0.1,0.2,...,1},
                yticklabel style={
                    /pgf/number format/.cd,
                        fixed,
                        fixed zerofill,
                        precision=1
                }
            },
            ticklabel style={
                /pgf/number format/.cd,
                    fixed,
                    fixed zerofill,
                    precision=1
            },
            title={{helium retention} [\%]},
            xlabel={model output},
        ]
        \addplot graphics[xmin=0, xmax=8, ymin=0, ymax=8] {plt/pisces/ftx/parity_he_retention.png};
        \addplot[line] coordinates {(0, 0) (8, 8)};
        \node[label, yshift=-1pt] at (rel axis cs:0, 1) {$e_{\mathrm{train}} = 0.266$};
        \node[label, yshift=-11pt] at (rel axis cs:0, 1) {$e_{\mathrm{test}} = 0.293$};
        \node[label, yshift=-21pt] at (rel axis cs:0, 1) {$|u| = 1$};
        \nextgroupplot[%
            legend style={%
                at={(1.05, 1.04)},
                font={\normalsize},
                anchor=north west,
                /tikz/every odd column/.append style={column sep=1mm}
            },
            clip mode=individual,
            xlabel={time [\si{\second}]},
            ylabel={total sensitivity index},
            xmin=0, xmax=1,
            ymin=1e-2, ymax=1,
            ymode=log,
            ylabel shift=-5pt,
        ]
        \plotsens{17}{SkyBlue}{dat/pisces/ftx/gsa_surface_growth}{}
        \addlegendentry{\strut$E_s$}
        \plotsens{16}{Plum}{dat/pisces/ftx/gsa_surface_growth}{}
        \addlegendentry{\strut$E_{\textrm{in}}$}
        \plotsens{15}{Red}{dat/pisces/ftx/gsa_surface_growth}{}
        \addlegendentry{\strut$E_{\ce{He}_1}$}
        \plotsens{14}{SeaGreen}{dat/pisces/ftx/gsa_surface_growth}{}
        \addlegendentry{\strut$E_{\ce{V}_1}$}
        \plotsens{13}{RedViolet}{dat/pisces/ftx/gsa_surface_growth}{}
        \addlegendentry{\strut$E_{f, \ce{W}}$}
        \plotsens{12}{Rhodamine}{dat/pisces/ftx/gsa_surface_growth}{}
        \addlegendentry{\strut$E_{f, \ce{He}}$}
        \plotsens{11}{Aquamarine}{dat/pisces/ftx/gsa_surface_growth}{}
        \addlegendentry{\strut$a_0$}
        \node[] at (rel axis cs:0.3, 0.85) {$E_s$};
        \node[] at (rel axis cs:0.6, 0.4) {$E_{\textrm{in}}$};
        \node[] at (rel axis cs:0.125, 0.16) {$E_{\ce{He}_1}$};
        \node[] at (rel axis cs:0.35, 0.05) {$E_{\ce{V}_1}$};
        \addplot[error, smooth] table[x index=0, y index=1] {dat/pisces/ftx/error-1.dat};
        \nextgroupplot[%
            legend style={%
                at={(1.05, 1.04)},
                font={\normalsize},
                anchor=north west,
                /tikz/every odd column/.append style={column sep=1mm}
            },
            clip mode=individual,
            xlabel={time [\si{\second}]},
            xmin=0, xmax=1,
            ymin=1e-2, ymax=1,
            ymode=log,
            yticklabel=\empty
        ]
        \plotsens{17}{Red}{dat/pisces/ftx/gsa_he_retention}{}
        \addlegendentry{\strut$E_{\ce{He}_1}$}
        \plotsens{16}{Plum}{dat/pisces/ftx/gsa_he_retention}{}
        \addlegendentry{\strut$E_{\textrm{in}}$}
        \plotsens{15}{NavyBlue}{dat/pisces/ftx/gsa_he_retention}{}
        \addlegendentry{\strut$E_{\ce{He}_2}$}
        \plotsens{14}{SkyBlue}{dat/pisces/ftx/gsa_he_retention}{}
        \addlegendentry{\strut$E_s$}
        \plotsens{13}{Lavender}{dat/pisces/ftx/gsa_he_retention}{}
        \addlegendentry{\strut$E_{\ce{He}_4}$}
        \plotsens{12}{BlueViolet}{dat/pisces/ftx/gsa_he_retention}{}
        \addlegendentry{\strut$b_{\ce{I}_l}$}
        \plotsens{11}{Gray}{dat/pisces/ftx/gsa_he_retention}{}
        \addlegendentry{\strut$E_{\ce{He}_5}$}
        \plotsens{10}{Mahogany}{dat/pisces/ftx/gsa_he_retention}{}
        \addlegendentry{\strut$C_{\ce{V}_1}$}
        \plotsens{9}{SeaGreen}{dat/pisces/ftx/gsa_he_retention}{}
        \addlegendentry{\strut$E_{\ce{V}_1}$}
        \node[] at (rel axis cs:0.3, 0.8) {$E_{\ce{He}_1}$};
        \node[] at (rel axis cs:0.8, 0.475) {$E_{\textrm{in}}$};
        \node[] at (rel axis cs:0.625, 0.25) {$E_s$};
        \node[] at (rel axis cs:0.5, 0.075) {$E_{\ce{He}_4}$};
        \addplot[error, smooth] table[x index=0, y index=1] {dat/pisces/ftx/error-2.dat};
    \end{groupplot}
    \node (title) at ($(group c1r1.center)!0.5!(group c2r1.center)+(0,3cm)$) {\FTX{} \textbf{(17 parameters),} \PISCES{} \textbf{setting}};
\end{tikzpicture}
	\tikzexternaldisable

    \caption{Comparison of the predicted outputs from the surrogate and the actual model outputs (\emph{top}) and total sensitivity indices as a function of time (\emph{bottom}) for surface growth (\emph{left}) and helium retention (\emph{right}) for \FTX{} in the \PISCES{} setting.}
    \label{fig:pisces_ftx}
\end{figure*}

In the top row of \cref{fig:pisces_ftx}, we show a comparison between the actual \FTX{} outputs and the PCE surrogate predictions. Notice how the surrogate models are now much less accurate compared to the surrogates for the coupled \FTX{} code in the \ITERHe{} setting. However, we deem the accuracy of the surrogates to be sufficient in order to identify at least the most important parameters. This inferior accuracy is probably due to the larger dimensionality of the input space (17 instead of 15) and the limited computational budget to obtain the training samples, as well as the increase in the number of bursting events compared to the \ITERHe{} setting. Below, we will illustrate how a more accurate surrogate model can be constructed with a reduced number of parameters.

The corresponding sensitivity indices for each parameter are reported in the bottom row of \cref{fig:pisces_ftx}. Again, the solid line is a measure of the reliability of the sensitivity indices in the presence of surrogate error. For the surface growth, the most important parameters, sorted according to the sum of their sensitivity indices across time, are the surface binding energy of tungsten ($E_s$), the ion impact energy ($E_{\textrm{in}}$) and the migration energy $E_{\ce{He}_1}$. For the helium retention, the most important parameters are the migration energy $E_{\ce{He}_1}$, the ion impact energy ($E_{\textrm{in}}$) and the surface binding energy of tungsten ($E_s$).

In order to explain 99\% of the output variability across all time instances, it is sufficient to include those three parameters, together with the migration energy $E_{\ce{V}_1}$. With this key set of parameters identified, we construct a more accurate surrogate model using only this subset of parameters. This surrogate can then be used to replace the expensive evaluation of the coupled \FTX{} code in the multiscale framework of~\cite{lasa2021}. To this end, we also include the ion impact angle $\alpha_{\textrm{in}}$ to the reduced set of parameters, because it is considered to be an important operating condition.

\begin{figure*}[t]
    \centering
	\tikzsetnextfilename{pisces_ftx_reduced}
	\tikzexternalenable
	\setlength{\figurewidth}{7cm}
\setlength{\figureheight}{5.75cm}
\begin{tikzpicture}[%
        line/.style={draw, black, dashed, line cap=round},
        label/.style={anchor=north west, font=\scriptsize, xshift=1pt},,
        error/.style={default line, black}
    ]
    \begin{groupplot}[%
            group style={%
                group size=2 by 2,
                horizontal sep=5em,
                vertical sep=3em,
                xlabels at=edge bottom,
                ylabels at=edge left,
            },
            default axis,
            axis on top,
            legend style={%
                at={(1, 1)},
                anchor=north east,
                /tikz/every odd column/.append style={column sep=1mm}
            },
            clip mode=individual,
            ylabel={PCE surrogate output},
            title style={%
                align=center,
                yshift=-5pt,
                font=\small
            },
            ylabel style={at={(-0.15,0.5)}, anchor=south}
        ]
        \nextgroupplot[%
            xmin=-5, xmax=4,
            ymin=-5, ymax=4,
            xtick={-5, -2, ..., 4},
            ytick={-5, -2, ..., 4},
            title={{surface growth} [\si{\nano\metre}]},
            xlabel={model output},
	    ]
        \addplot graphics[xmin=-5, xmax=4, ymin=-5, ymax=4] {plt/pisces/ftx-reduced/parity_surface_growth.png};
        \addplot[line] coordinates {(-5, -5) (4, 4)};
        \node[label, yshift=-1pt] at (rel axis cs:0, 1) {$e_{\mathrm{train}} = 0.151$};
        \node[label, yshift=-11pt] at (rel axis cs:0, 1) {$e_{\mathrm{test}} = 0.172$};
        \node[label, yshift=-21pt] at (rel axis cs:0, 1) {$|u| = 3$};
        \nextgroupplot[%
            xmin=0, xmax=8,
            ymin=0, ymax=8,
            title={{helium retention} [\%]},
            xlabel={model output},
            colormap/jet,
            colorbar,
            colorbar style={%
                ylabel={time [\si{\second}]},
                ylabel style={at={(2.5,0.5)}, anchor=north},
                ymin=0.1, ymax=1,
                point meta min=0.1, point meta max=1,
                width=1em,
                at={(1.05, 0.5)},
                anchor=west,
                ytick style={draw=none},
                ytick={0.1,0.2,...,1},
                yticklabel style={
                    /pgf/number format/.cd,
                        fixed,
                        fixed zerofill,
                        precision=1
                }
            },
	    ]
        \addplot graphics[xmin=0, xmax=8, ymin=0, ymax=8] {plt/pisces/ftx-reduced/parity_he_retention.png};
        \addplot[line] coordinates {(0, 0) (8, 8)};
        \node[label, yshift=-1pt] at (rel axis cs:0, 1) {$e_{\mathrm{train}} = 0.084$};
        \node[label, yshift=-11pt] at (rel axis cs:0, 1) {$e_{\mathrm{test}} = 0.085$};
        \node[label, yshift=-21pt] at (rel axis cs:0, 1) {$|u| = 3$};
        \nextgroupplot[%
            legend style={%
                at={(1.05, 1.04)},
                font={\normalsize},
                anchor=north west,
                /tikz/every odd column/.append style={column sep=1mm}
            },
            clip mode=individual,
            xlabel={time [\si{\second}]},
            ylabel={total sensitivity index},
            xmin=0, xmax=1,
            ymin=1e-2, ymax=1,
            ymode=log,
            ylabel shift=-5pt,
        ]
        \plotsens{5}{SkyBlue}{dat/pisces/ftx-reduced/gsa_surface_growth}{}
        \addlegendentry{\strut$E_s$}
        \plotsens{4}{Plum}{dat/pisces/ftx-reduced/gsa_surface_growth}{}
        \addlegendentry{\strut$E_{\textrm{in}}$}
        \plotsens{3}{Red}{dat/pisces/ftx-reduced/gsa_surface_growth}{}
        \addlegendentry{\strut$E_{\ce{He}_1}$}
        \plotsens{2}{LimeGreen}{dat/pisces/ftx-reduced/gsa_surface_growth}{}
        \addlegendentry{\strut$\alpha_{\textrm{in}}$}
        \node[] at (rel axis cs:0.2, 0.875) {$E_s$};
        \node[] at (rel axis cs:0.8, 0.625) {$E_{\textrm{in}}$};
        \node[] at (rel axis cs:0.8, 0.45) {$E_{\ce{He}_1}$};
        \node[] at (rel axis cs:0.8, 0.2) {$\alpha_{\textrm{in}}$};
        \addplot[error, smooth] table[x index=0, y index=1] {dat/pisces/ftx-reduced/error-1.dat};
        \nextgroupplot[%
            legend style={%
                at={(1.05, 1.04)},
                font={\normalsize},
                anchor=north west,
                /tikz/every odd column/.append style={column sep=1mm}
            },
            clip mode=individual,
            xlabel={time [\si{\second}]},
            xmin=0, xmax=1,
            ymin=1e-2, ymax=1,
            ymode=log,
            yticklabel=\empty
        ]
        \plotsens{5}{Red}{dat/pisces/ftx-reduced/gsa_he_retention}{}
        \addlegendentry{\strut$E_{\ce{He}_1}$}
        \plotsens{4}{Plum}{dat/pisces/ftx-reduced/gsa_he_retention}{}
        \addlegendentry{\strut$E_{\textrm{in}}$}
        \plotsens{3}{SkyBlue}{dat/pisces/ftx-reduced/gsa_he_retention}{}
        \addlegendentry{\strut$E_s$}
        \plotsens{2}{LimeGreen}{dat/pisces/ftx-reduced/gsa_he_retention}{}
        \addlegendentry{\strut$\alpha_{\textrm{in}}$}
        \node[] at (rel axis cs:0.3, 0.7) {$E_{\ce{He}_1}$};
        \node[] at (rel axis cs:0.875, 0.4) {$E_{\textrm{in}}$};
        \node[] at (rel axis cs:0.45, 0.115) {$E_s$};
        \addplot[error, smooth] table[x index=0, y index=1] {dat/pisces/ftx-reduced/error-2.dat};
    \end{groupplot}
    \node (title) at ($(group c1r1.center)!0.5!(group c2r1.center)+(0,3cm)$) {\FTX{} \textbf{(5 parameters),} \PISCES{} \textbf{setting}};
\end{tikzpicture}
	\tikzexternaldisable

    \caption{Comparison of the predicted outputs from the surrogate and the actual model outputs (\emph{top}) and total sensitivity indices as a function of time (\emph{bottom}) for surface growth (\emph{left}) and helium retention (\emph{right}) for \FTX{} in the \PISCES{} setting using only the 5 most important parameters.}
    \label{fig:pisces_ftx_reduced}
\end{figure*}
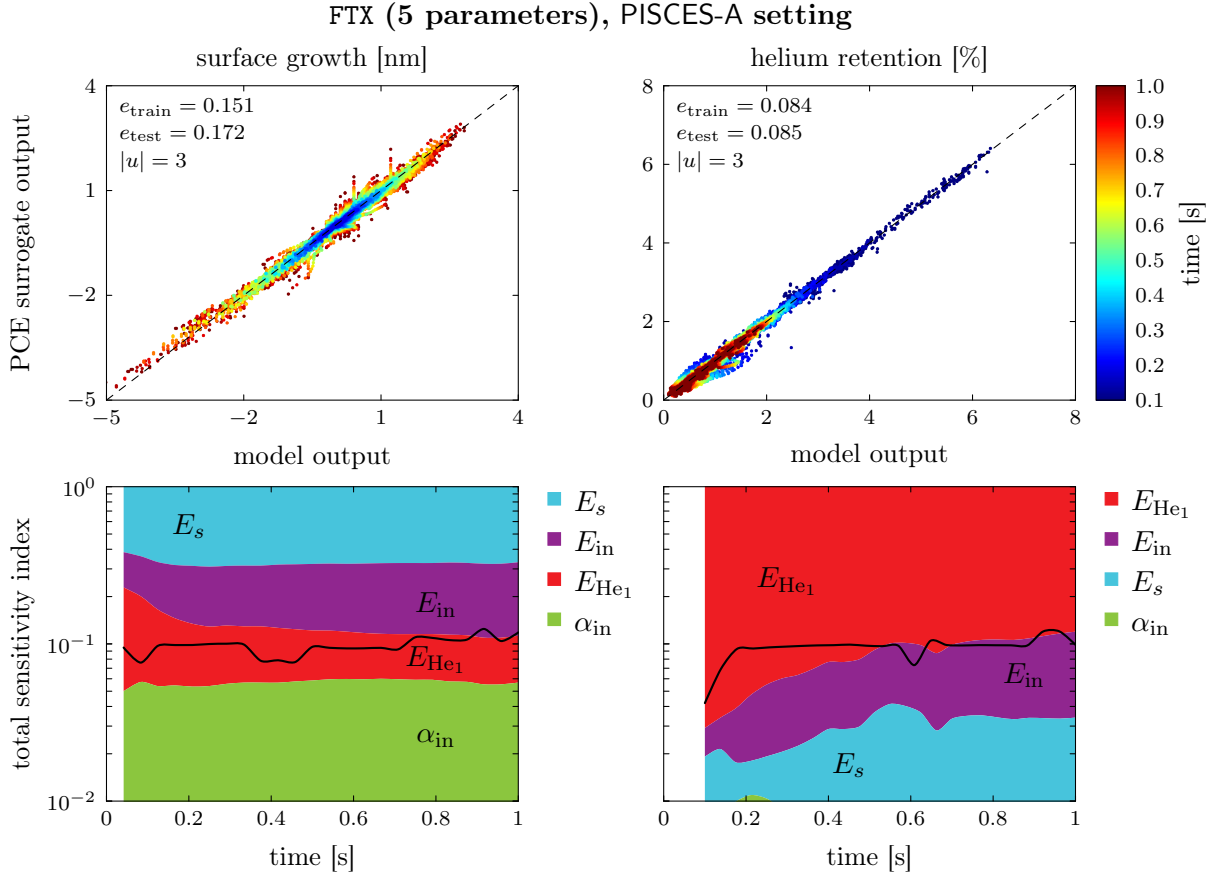

A comparison between the actual \FTX{} outputs and the surrogate predictions including only the 5 most important parameters is shown in the top row of~\cref{fig:pisces_ftx_reduced}. Note that the surrogate models are now much more accurate, reducing the test errors $e_{\textrm{test}}$ from 0.427 to 0.172 for the surface growth, and from 0.293 to 0.085 for the helium retention. The corresponding total sensitivity indices as a function of time are shown in the bottom row of~\cref{fig:pisces_ftx_reduced}. Again, the solid line in this plot is a measure of the reliability of the sensitivity indices in the presence of surrogate error. Note that, for the helium retention, the predicted sensitivity indices for the migration energy $E_{\ce{He}_1}$, the ion impact energy $E_{\textrm{in}}$, and the surface binding energy of tungsten $E_s$, agree with the ones obtained from the 17-dimensional surrogate in terms of their relative magnitudes. For the surface growth, the migration energy $E_{\ce{He}_1}$ has been attributed a larger share of the total output variance, compared to the predicted sensitivities from the 17-dimensional surrogate model. Also, the ion impact angle ($\alpha_{\textrm{in}}$) seems to be more important than the migration energy $E_{\ce{V}_1}$, something we did not observe with the 17-dimensional surrogate model. Note, however, that the sensitivity index for the ion impact angle is below the computed threshold reflecting surrogate accuracy, and its apparent importance may be due to the error in the surrogate.

\subsection{Discussion}\label{sec:discussion}

One of the insights gained from the GSA study for \FTridyn{} is the importance of the IEAD. For example, in the \PISCES{} case discussed in \cref{sec:gsa_study_for_the_pisces_setting}, where ion impact energy $E_\textrm{in}$ and angle $\alpha_\textrm{in}$ are considered as uncertain parameters, these parameters are consistently identified as the most important parameters associated with \FTridyn{}. The ion impact energy $E_\textrm{in}$ seems to be more important than the ion impact angle $\alpha_\textrm{in}$, probably because of the small range (\SI{0}{\degree} -- \SI{30}{\degree}) used during the study, see \cref{tab:uncertain_parameters}. In effect, the sputtering yield reaches its maximum around \SI{60}{\degree} to \SI{70}{\degree} under this target-projectile combination, see~\cite{behrisch2007}, a value which is unrealistic in the \PISCES{} case. On the other hand, in the \ITERHe{} setting considered in \cref{sec:gsa_study_for_the_iter_he_setting}, the IEAD is provided by \texttt{hPIC}, another code in the multiscale code hierarchy, in the form of a distribution in the energy-angle space, see, e.g.,~\cite{lasa2020}. This is in contrast to the mono-energetic, mono-angle irradiation considered in the \PISCES{} setting. In order to keep the number of input parameters tractable, we did not include the energy-angle distribution in the UQ study, but instead chose to consider the IEAD as a fixed, deterministic quantity. Under such conditions, the surface binding energy of tungsten was identified as the most important parameter. Including the IEAD in the GSA study will require a study of coupled \texttt{hPIC}-\texttt{FTX}. A code coupling effort for the latter is currently ongoing.

The migration energy of the \ce{He1} cluster, $E_{\ce{He}_1}$, is consistently identified as the most important parameter in predicting the helium retention, for both the \PISCES{} and \ITERHe{} settings we considered in this study. This makes sense, given that it is the most abundant and mobile species. We should also note that the migration energy of self-interstitials is not subject to uncertainty, because they are assumed to diffuse much faster than the other clusters.

Furthermore, we found that the importance of the migration energy of the \ce{He1} cluster, $E_{\ce{He}_1}$, diminishes over time, see \cref{fig:iter_he_ftx,fig:pisces_ftx}. A possible explanation could be that, as more clusters are formed, the tungsten lattice contains more sinks where the newly generated, mobile \ce{He1} clusters can get trapped. Initially, in the pristine tungsten lattice, the \ce{He1} clusters migrate over a longer period of time, until they eventually diffuse back to the surface and desorb, self-cluster or join an existing, small cluster. As more and more clusters appear over time, especially in the near-surface, the implanted \ce{He1} is shorter-lived, and hence its importance diminishes over time.

Considering the coupled \FTX{} setup for the \PISCES{} setting, the GSA results presented in \cref{fig:iter_he_ftx,fig:pisces_ftx} reveal that the surface binding energy of tungsten and the ion impact energy, two input parameters of \FTridyn{}, appear to be important for both the surface growth and the helium retention, through their effect on the sputtering yield. This observation confirms the advantage of considering the coupled setting, as such coupling is clearly crucial to the system behavior.

Also from \cref{fig:iter_he_ftx,fig:pisces_ftx}, the surface binding energy of tungsten only seems to be important in the \PISCES{} case, and not in the \ITERHe{} setting. In fact, none of the \FTridyn{} input parameters appear to be important for the \ITERHe{} case. A possible explanation for this observation is as follows. The sputtering yield controls how quickly the near-surface \ce{He}-containing layer is removed. The surface binding energy determines how strongly a tungsten atom is bound to the surface, and directly affects the energy threshold required for a sputtering event, i.e., it determines the amount of energy an incoming ion needs to transfer to the surface atom to sputter it. If the energy of the incoming flux is larger than the energy threshold required for sputtering, as is the case in the \PISCES{} setting, changes in the surface binding energy will be important. However, when the energy of the incoming flux is much lower than the energy threshold required for sputtering, as is the case in the \ITERHe{} setting, changes in the surface binding energy will have little to no effect. Besides the secondary effect caused by changes in the sputtering yield, the ion impact energy also directly affects the implantation depth, which can impact the helium retention because helium implanted deeper in the substrate is less likely to undergo trap mutation and become immobile.

Furthermore, it appears that the migration energy of the \ce{He1} clusters $E_{\ce{He}_1}$, which was identified as the most important parameter for the helium retention, has little or no effect on the surface growth for the \PISCES{} settings, in contrast to the \ITERHe{} experiments, see \cref{fig:pisces_ftx}. Again, a possible explanation may be the larger value for the incoming flux in the \PISCES{} experiments compared to the relatively low plasma density anticipated for early \ITERHe{} operation with a helium plasma. Only parameters that introduce a change in the trap mutation may impact the surface growth. In particular, in order for the surface to grow, we require a helium clustering event creating a vacancy and an interstitial. The latter interstitial may eventually diffuse towards the surface, leading to surface growth. Under a sufficiently large incoming helium flux, such as the one used in the \PISCES{} setting, the implanted helium experiences a higher rate of self-clustering, or joining other clusters, relative to the timescale where the diffusion, determined by the migration energies, matters. As such, we expect the migration energies to be less important under high-flux conditions, such as the \PISCES{} setting.

\section{Conclusions and Future Work}\label{sec:conclusions_and_future_work}

Although there have been various attempts to model the mechanics involved in the material subsurface evolution of tungsten exposed to helium plasmas, a global sensitivity analysis framework to investigate the effect of model parameters in these models has hitherto been lacking. Yet, information about the sensitivity of model outputs to these input parameters is of crucial importance in order to improve understanding of the underlying complex physical processes and interactions that control the subsurface morphology, as well as to facilitate further model development, and ultimately to target new experiments that can be used to reduce the model uncertainty. As the computer codes used to simulate these models require several hours to days of computation time on HPC systems for even a single model evaluation, it is important to use efficient sensitivity analysis methods that limit the size of the ensemble of model evaluations, but provide sufficient accuracy to extract the needed information. In this work, we employed such a global sensitivity analysis framework using polynomial chaos surrogate models that allow extraction of the required sensitivity information directly from the polynomial coefficients. These sparse, high-dimensional nonlinear polynomial surrogates are constructed by relying on Bayesian compressive sensing methods~\cite{sargsyan2014}. 

We used our framework to perform a global sensitivity analysis of \FTX{}, a coupled code part of a larger multiscale, multiphysics modeling attempt to predict the material evolution of plasma-facing components in future fusion reactors~\cite{lasa2021}. We considered two problem settings: the \ITERHe{} setting that mimics ITER-like conditions of early operation with a helium plasma, and the \PISCES{} setting that mimics the experimental conditions inside the PISCES-A linear plasma device. Given specified uncertainty in model parameters/inputs, our analysis highlights the importance of individual parameters as measured by their fractional contribution to the variance in uncertain model outputs of interest. Our results indicate that, for the \ITERHe{} setting, where we assume the ion energy-angle distribution (IEAD) is fixed, 99\% of the variability in the surface growth and helium retention can be explained by the variation in the migration energies $E_{\ce{He}_1}$, $E_{\ce{He}_2}$ and $E_{\ce{He}_4}$. In the \PISCES{} setting, the most important parameters are the surface binding energy of tungsten, the ion impact energy and the migration energy $E_{\ce{He}_1}$. With this key set of important parameters identified, we constructed a more accurate surrogate model in only 5 dimensions. This surrogate model can be used to replace the coupled \FTX{} code in the multiphysics modeling hierarchy from~\cite{lasa2020}. Model calibration may also be performed with this more accurate surrogate model, using the data reported in, e.g.,~\cite{woller2015}. This will be the topic of future research efforts.

In addition, we plan to use our framework to repeat the global sensitivity analysis study for the coupled \FTX{} code, employing a recently developed non-stochastic bubble bursting model by the authors of~\cite{blondel2017a}. This new model should reduce the computational burden involved with running the coupled code, allowing larger ensemble sizes to be considered in the sensitivity analysis study. Finally, we mention that the more modern \texttt{RustBCA} code developed in~\cite{drobny2021} can be used as an alternative to the \FTridyn{} code we employed in this study. However, a code coupling between \texttt{RustBCA} and \Xolotl{} similar to~\cite{lasa2020} is yet to be developed.

Finally, the UQ framework developed in this paper will be used to perform global sensitivity analysis and surrogate construction for a larger part of the multiphysics framework modeling plasma and surface material interactions under construction\cite{lasa2020}. In particular, ongoing efforts include a sensitivity analysis for the coupling of the impurity transport code \texttt{GITR}, see~\cite{younkin2021}, with the ion-solid interaction code \FTridyn{}, and for the coupled \texttt{GITR}--\FTX{} simulation framework.

\section*{Acknowledgements}

This work was supported by the U.S.\,Department of Energy, Office of Science, Office of Fusion Energy Sciences and Office of Advanced Scientific Computing Research through the Scientific Discovery through Advanced Computing project on Plasma-Surface Interactions.

This research used resources of the National Energy Research Scientific Computing Center (NERSC), a U.S.\,Department of Energy Office of Science User Facility operated under Contract DE-AC02-05CH11231.

This article has been co-authored by employees of National Technology \& Engineering Solutions of Sandia, LLC under Contract No. DE-NA0003525 with the U.S. Department of Energy (DOE). The employees co-own right, title and interest in and to the article and are responsible for its contents. The United States Government retains and the publisher, by accepting the article for publication, acknowledges that the United States Government retains a non-exclusive, paid-up, irrevocable, world-wide license to publish or reproduce the published form of this article or allow others to do so, for United States Government purposes. The DOE will provide public access to these results of federally sponsored research in accordance with the DOE Public Access Plan https://www.energy.gov/downloads/doe-public-access-plan. 

\bibliographystyle{abbrv}
\bibliography{ref/references}

\begin{appendices}
\section{Bayesian compressive sensing}\label{sec:bayeseian_compressive_sensing}

The least-squares regression problem from \cref{eq:least_squares} can be written in matrix form as
\begin{align}
    \bsc^{\mathrm{LS}} &= \argmin_\bsc \|\bsy - \bsP \bsc\|_2^2, \label{eq:least_squares_matrix}
\end{align}
where $\bsy$ is an $N$-dimensional vector of model outputs with $y_n = y^{(n)}$ for $n = 1, 2, \ldots, N$, $\bsP$ is a matrix of size $N \times K$ with $P_{nk} = \Phi_{k}(\bsxi^{(n)})$ for $n = 1, 2, \ldots, N$ and $k = 1, 2, \ldots, K$, where $N$ is the number of training samples, $K$ is the number of PCE basis terms, and where $\|\cdot\|_2$ denotes the usual (Euclidean) $\ell_2$-norm.

When the number of model evaluations $N$ is lower than the number of polynomial terms $K$, the least-squares problem in~\eqref{eq:least_squares_matrix} is underdetermined. In such cases, the solution $\bsc$ is not well-defined, and one can add an $\ell_1$ regularization term that enforces sparsity in the polynomial expansion coefficients, i.e.,
\begin{equation}\label{eq:regularized_least_squares}
    \bsc^{\mathrm{CS}} = \argmin_\bsc \left(\|\bsq - \bsP \bsc\|^2_2 + \gamma \|\bsc\|_1\right),
\end{equation}
where $\gamma>0$ is a regularization parameter, typically determined using cross-validation, and where $\|\cdot\|_1$ denotes the $\ell_1$-norm. The regularization term will ensure that only a small subset of basis terms will be retained in the solution vector $\bsc^{\mathrm{CS}}$. This approach is in line with the compressive sensing (CS) method commonly used in signal recognition, see \cite{candes2006,donoho2006}.

\begin{algorithm*}
    \begin{algorithmic}[1]
    \Statex \strut\textbf{input:} training data $\mathscr{D}_{\textrm{train}}$, number of iterations $L$
    \Statex \textbf{output:} higher-order sparse basis $\calI_d^{(L)}$, coefficient vector $\bsc^{(L)}$
    \Statex
    \Procedure{\textsf{construct\texttt{\_}basis}}{$\mathscr{D}_{\textrm{train}}$, $L$}
    \State $\calI_d^{(0)} \gets \{\bszero\}$ \Comment{initialize the index set}
    \For{$\ell$ \textbf{from} 1 \textbf{to} $L$} \Comment{perform $L$ iterations}
    \State $\calI_\mathrm{new} \gets \calI_d^{(\ell - 1)}$
    \ForEach {$\fraku \in \calI_d^{(\ell - 1)}$} \Comment{loop over each index in the index set}
    \For{$j$ \textbf{from} 1 \textbf{to} $s$}
    \State $\calI_\mathrm{new} \gets \calI_\mathrm{new} \cup \{\fraku + \bse_j\}$ \Comment{add all forward neighbors}
    \EndFor
    \EndFor
    \State $\bsc^{(\ell)}, \calI_d^{(\ell)} \gets \textsf{BCS}(\mathscr{D}_{\textrm{train}}, \calI_\mathrm{new})$ \Comment{run BCS to obtain sparse basis}
    \EndFor
    \EndProcedure
    \end{algorithmic}
    \caption{Adaptive basis growth procedure from~\cite{sargsyan2014}.}
    \label{alg:wibcs}
\end{algorithm*}

In Bayesian compressive sensing (BCS), the regression problem in~\eqref{eq:regularized_least_squares} is embedded in a Bayesian framework, see, e.g.,~\cite{ji2008,babacan2009,sargsyan2014}. In BCS, one constructs an \emph{uncertain} PCE surrogate model by assigning a prior probability distribution $p(\bsc)$ to the coefficient vector $\bsc$. The goal is then to obtain the posterior distribution $p(\bsc | \mathscr{D}_\textrm{train})$, i.e., the probability distribution of the coefficient vector $\bsc$ after observing the training data $\mathscr{D}_\textrm{train}$. The posterior distribution is related to the prior probability via the likelihood function $\calL_{\mathscr{D}_\textrm{train}}(\bsc)$, a consequence of Bayes' law, i.e.,
\begin{equation}
    p(\bsc | \mathscr{D}_\textrm{train}) \propto \calL_{\mathscr{D}_\textrm{train}}(\bsc) p(\bsc).
\end{equation}
The likelihood expresses the goodness-of-fit of the PCE surrogate model with coefficients $\bsc$ to the available input-output data $\mathscr{D}_\textrm{train}$. Maximizing the (log of the) posterior probability, the optimal value for the coefficients is
\begin{equation}\label{eq:bayesian_compressive_sensing}
    \bsc^{\mathrm{BCS}} = \argmax_\bsc \left(\log \calL_{\mathscr{D}_\textrm{train}}(\bsc) + \log p(\bsc)\right).
\end{equation}
This is clearly equivalent to the CS formulation in~\eqref{eq:regularized_least_squares}, where the objective function is the negative log-likelihood and the regularization term is the negative log-prior. In particular, combining an i.i.d.\,Gaussian likelihood with constant variance $\sigma^2$, i.e.,
\begin{align}
    &\calL_{\mathscr{D}_\textrm{train}}(\bsc) = \\
    &\frac{1}{(\sqrt{2\pi}\sigma)^N} \exp \left( - \frac{1}{2\sigma^2} \sum_{n=1}^{N} \left(f(\bsx^{(n)}) - \sum_{k=1}^K c_k \Phi_k(\bsxi^{(n)})\right)^2 \right)
\end{align}
with a weighted Laplace prior
\begin{equation}
p(\bsc) = \left(\frac{\gamma}{2}\right)^{K} \exp\left(- \gamma \sum_{k=1}^K |c_k|\right),
\end{equation}
the solution of~\eqref{eq:regularized_least_squares} can be recovered, see, e.g.,~\cite{sargsyan2014}. A key strength of the BCS approach in \cref{eq:bayesian_compressive_sensing}, however, is that, when only a small number of training samples are available, as will be the case in our numerical experiments later on, it leads to a probabilistic surrogate model that quantifies the additional uncertainty in the prediction through the posterior distribution $p(\bsc | \mathscr{D}_\textrm{train})$ of the PCE coefficients.

It can be shown that the BCS solution for the coefficients $\bsc^{\mathrm{BCS}}$ is a multivariate Gaussian posterior distribution, and a fast approximate solution can be obtained using techniques borrowed from the relevance vector machine (RVM) literature, see, e.g.,~\cite{tipping2003}.

An adaptive basis growth procedure to construct the PCE index set $\calI_d$ has been suggested in~\cite{sargsyan2014}. Starting from an initial set of basis terms $\calI_d^{(0)} = \{\bszero\}$, we gradually enrich the polynomial basis $\calI_d^{(\ell)}$ for $\ell = 1, 2, \ldots, L$ with more terms, in each step running the BCS algorithm to reduce the set of basis vectors and only retain the relevant components. The full procedure is shown in~\cref{alg:wibcs}. The procedure is terminated when a predefined maximum number of iterations $L$ is reached. Alternatively, the procedure may be terminated early when there is no change in the index set $\calI_d^{(\ell)}$ for two successive iterations $\ell - 1$ and $\ell$. In practice,~\cref{alg:wibcs} is repeated $R=3$ times, each time using a random partitioning of the data $\mathscr{D}$ into 90\% training samples and 10\% test samples, to avoid overfitting. The final index set is then computed as the intersection of the $R$ trial index sets $\calI_d^{(r, L)}$, to avoid any remaining overfitting, i.e.,
\begin{equation}
    \calI_d = \bigcap_{r=1}^R \calI_d^{(r, L)}.
\end{equation}
Once the final index set $\calI_d$ has been obtained, we perform a final least-squares fit to compute the coefficients of the PCE with only the remaining basis terms.
\end{appendices}

\end{document}